\documentclass{ws-ijmpd}
\usepackage[super,compress]{cite}
\usepackage{color}
\usepackage{hyperref}

\newcommand{\bq}{\begin{equation}}
\newcommand{\eq}{\end{equation}}
\newcommand{\bqa}{\begin{eqnarray}}
\newcommand{\eqa}{\end{eqnarray}}
\def\ba#1\ea{\begin{align}#1\end{align}}

\newcommand{\Om}{\Omega_{\rm m}}
\newcommand{\rhom}{\rho_{\rm m}}
\newcommand{\rhomb}{\bar{\rho}_{\rm m}}
\newcommand{\drhom}{\delta\rho_{\rm m}}

\newcommand{\Vm}{V_{\rm m}}
\newcommand{\Deltam}{\Delta_{\rm m}}

\newcommand{\RTH}{r_{\rm th}}
\newcommand{\yhal}{y_{\rm h}}
\newcommand{\yenv}{y_{\rm env}}
\newcommand{\deltac}{\delta_{\rm c}}

\newcommand{\Geff}{G_{\rm eff}}

\newcommand{\delg}{\delta g}
\newcommand{\Mbar}{\bar{M}}
\newcommand{\Sm}{S_{\rm m}}
\newcommand{\aK}{\alpha_{\rm K}}
\newcommand{\aM}{\alpha_{\rm M}}
\newcommand{\aB}{\alpha_{\rm B}}
\newcommand{\aT}{\alpha_{\rm T}}

\newcommand{\cs}{c_{\rm s}}

\def\lsim{\, \lower .75ex \hbox{$\sim$} \llap{\raise .27ex \hbox{$<$}} \,}

\begin{document}

\markboth{Lucas Lombriser}
{Parametrizations}

%
\catchline{}{}{}{}{}
%

\title{PARAMETRIZATIONS FOR TESTS OF GRAVITY}

\author{LUCAS LOMBRISER}

\address{D\'epartement de Physique Th\'eorique, 
Universit\'e de Gen\`eve, 24 quai Ansermet, \\ CH-1211~Gen\`eve~4, Switzerland\\
Institute for Astronomy, University of Edinburgh, Royal Observatory, Blackford Hill, \\ Edinburgh, EH9 3HJ, United Kingdom\\
lucas.lombriser@unige.ch}

\maketitle

\begin{history}
\received{Day Month Year}
\revised{Day Month Year}
\end{history}

\begin{abstract}
With the increasing wealth of high-quality astronomical and cosmological 
data and the manifold departures from General Relativity in principle 
conceivable, the development of generalized parameterization frameworks that unify gravitational models and cover a wide range of length scales and a variety of observational probes to enable systematic high-precision tests of gravity has been a stimulus for intensive research.
A review is presented here for some of the formalisms devised for this purpose,
covering the cosmological large- and small-scale structure, the astronomical static weak-field regime as well as emission and propagation effects for gravitational waves.
This includes linear and nonlinear parametrized post-Friedmannian frameworks, effective field theory approaches, the parametrized post-Newtonian expansion, the parametrized post-Einsteinian formalism as well as an inspiral-merger-ringdown waveform model among others.
Connections between the different formalisms are highlighted where they have been established and a brief outlook is provided for general steps towards a unified global framework for tests of gravity and dark sector models.
\end{abstract}


\ccode{PACS numbers:}


\section{Introduction} \label{sec:intro}

With the great volume of modified gravity, dark energy, and dark sector interaction models that have been proposed as extensions to General Relativity (GR) and the Standard Model~\cite{Copeland:2006wr,Clifton2011,Joyce2014,Koyama2016,Bull2016,Joyce2016,Ishak2018}, the development of a more systematic approach to explore their astronomical and cosmological implications has been and continues to be of prominent interest.
The parametrized post-Newtonian (PPN)~\cite{Will1971,Will1993} expansion fulfills this purpose for
astrophysical phenomena
amenable to a low-energy static approximation,
and stringent model-independent constraints have been inferred using this formalism, particularly from employing observations in the Solar System~\cite{Will2014}.
The PPN
expansion, however, is not straightforwardly applicable to gravitational modifications endowed with screening mechanisms~\cite{Joyce2014,Joyce2016} and corresponding generalizations have only been developed more recently~\cite{AvilezLopez2015,McManus2017}.

More crucially, the formalism is not suitable for evolving backgrounds such as encountered in cosmology.
Inspired by the successes of PPN, the development of an equivalent counterpart formalism on cosmological scales has therefore been the objective of intensive research.
The frameworks devised for this purpose are generally referred to as parametrized post-Friedmannian (PPF) formalisms~\cite{Uzan2006,Caldwell2007,Zhang2007,Amendola2007,Hu2007}, a term that has been coined by Ref.~\refcite{Tegmark2001}, and particularly adopted for the
formalisms
of Refs.~\refcite{Hu2007} and~\refcite{Baker2012}.
They aim at providing a consistent generalized description for the gravitational effects
in the cosmological background and formation of structure.
At the level of linear cosmological perturbations, this comprises both purely phenomenological parametrizations~\cite{Peebles1980,Linder2005,Uzan2006,Caldwell2007,Zhang2007,Amendola2007,Hu2007} as well as theoretically more constrained effective field theory (EFT)~\cite{Creminelli2009,Park2010,Gubitosi2012,Bloomfield2012,Gleyzes2013,Tsujikawa2014,Bellini2014,Gleyzes2014,Lagos2017} approaches.

Eventually linear theory fails in describing the cosmological structure below a few tens of Mpc.
Together with the screening effects, this severely complicates performing generalized and consistent parametrized tests of gravity.
However, a number of approaches have been developed to
access the
nonlinear scales, including generalized cosmological perturbation theory~\cite{Koyama2009,Brax2012a,Bellini2015,Taruya2016,Bose2016,Yamauchi2017,Frusciante2017,Bose2018,Hirano2018}, interpolation functions bridging the modified and screened regimes~\cite{Hu2007,Zhao2011}, parametrizations of screening mechanisms combined with nonlinear structure-formation models~\cite{Lombriser2016b,Hu2017}, or a unified parametrization of the Lagrangian for a class of modified gravity models~\cite{Brax2012a}.

Further approaches to extending the PPN formalism to both linear and nonlinear cosmological scales have been pursued with patching together post-Newtonian expansions in small regions of spacetime to a parametrized post-Newtonian cosmology (PPNC)~\cite{Sanghai2017} or by performing a post-Friedmannian expansion of the cosmological metric in powers of the inverse light speed in vacuum~\cite{Milillo2015}, which however remains to be extended for parametrized gravitational modifications.

With the dawn of gravitational wave astronomy heralded by the LIGO and Virgo detections~\cite{Abbott2016a,Abbott2017}, powerful new tests of gravity are being facilitated.
Effects of modifying gravity can both manifest in the emitted waveforms by modifications at the source~\cite{Yunes2009,Li2011b,Mirshekari2013,Lang2014,Abbott2016,Abbott2016b} but also in the observed waveforms entering through effects on the propagation of the wave~\cite{Saltas2014,Lombriser2015c}.
Cosmological propagation effects are well described by the PPF or EFT formalisms whereas source modifications can be captured by a parametrized post-Einsteinian (ppE)~\cite{Yunes2009} framework and the generalized inspiral-merger-ringdown waveform model~\cite{Li2011b}.
The gravitational effects in the different formalisms can also be unified~\cite{Nishizawa2017}, where however an incorporation of screening mechanisms remains to be elaborated.

This {\bf article} reviews the different frameworks that have been developed to parametrize gravitational effects for different astrophysical probes and scales and discusses connections between the formalisms where they have been elaborated, also providing an outlook for how such connections may be established where they are currently missing.
Sec.~\ref{sec:ppf} first reviews the PPF framework.
It addresses the evolution of the cosmological background in Sec.~\ref{sec:background} with parametrizations of dark energy~\cite{Chevallier2001,Linder2003} and a classification of genuine self-acceleration from modifying gravity~\cite{Lombriser2015c,Lombriser2016a}.
Sec.~\ref{sec:linear} discusses parametrization frameworks on linear scales, ranging from the modified large-scale structure with the growth-index parametrization~\cite{Peebles1980,Linder2005}, closure relations for modified linear perturbations~\cite{{Uzan2006,Caldwell2007,Zhang2007,Amendola2007,Hu2007}}, and EFT to the modified gravitational wave propagation and the choice of time and scale dependence in the parametrized modifications.
Parametrizations of modified gravity effects on the nonlinear cosmological structure are reviewed in Sec.~\ref{sec:nonlinear}, separating the weekly and deeply nonlinear regimes.
Sec.~\ref{sec:PPN} is devoted to the PPN formalism, briefly reviewing the post-Newtonian expansion in Sec.~\ref{sec:PN}, the parametrization of modified gravity effects on this expansion in Sec.~\ref{sec:PPNexpansion}, and the incorporation of screening mechanisms in Sec.~\ref{sec:PPNscreening}.
The parametrization of gravitational waveforms is addressed in Sec.~\ref{sec:waveforms} and a brief overview of further parametrizations for tests of gravity is given in Sec.~\ref{sec:further}.
Finally, a summary of the {\bf article} and a brief outlook for the field is presented in Sec.~\ref{sec:discussion}.

The following conventions will be adopted: the metric signature is $-+++$, primes denote derivatives with respect to $\ln a$ unless indicated otherwise, scalar perturbations of the Friedmann-Lema\^itre-Robertson-Walker (FLRW) metric are defined in Newtonian gauge as $\Psi\equiv\delta g_{00}/(2g_{00})$ and $\Phi\equiv\delta g_{ii}/(2g_{ii})$, bold mathematical symbols denote spatial vectors, and the speed of light in vacuum is set to unity unless otherwise specified.

\section{Parametrized Post-Friedmannian Frameworks} \label{sec:ppf}

In order to enable generalized analyses and observational tests of the cosmological effects of modified gravity and dark energy, extensive efforts have been invested in the development of a cosmological counterpart formalism to the static PPN expansion.
The different PPF approaches
aim at consistently unifying the description of generalized modified gravity and dark energy effects in the cosmological background evolution as well as the linear and nonlinear cosmological structure formation.
A further connection to the cosmological effects on the propagation of gravitational waves has also become desirable with the increasing amount of gravitational wave data.

The background evolution in the PPF formalism is addressed first
in Sec.~\ref{sec:background}.
Sec.~\ref{sec:linear} then discusses the linear perturbation regime, both for large-scale structure and gravitational wave propagation.
Finally, recent progress towards an extension of the linear PPF formalism to the nonlinear regime of cosmological structure formation is reviewed in Sec.~{\ref{sec:nonlinear}.

\subsection{Background cosmology} \label{sec:background}

Modifications of gravity or dark energy models that differ substantially from a cosmological constant can affect the geometry of spacetime in an observable manner.
At the background level the two extensions to standard cosmology can however not be distinguished in general.
They therefore share a parametrization for their effects on the expansion history, often characterized by the dark energy equation of state $w(t)$.
Sec.~\ref{sec:w} briefly discusses the most common parametrization for $w(t)$.
Much of the motivation for modifications of gravity on cosmological scales has been drawn from offering an alternative explanation to dark energy or the cosmological constant as the driver for the observed late-time accelerated expansion of our Universe.
Sec.~\ref{sec:cosmicacceleration} provides a discussion of the state of cosmic self-acceleration due to genuine modifications of gravity from a parametrized perspective.

\subsubsection{Dark energy equation of state} \label{sec:w}

The evolution of a spatially-flat homogeneous and isotropic cosmological background
is to full generality described by only one free function of time, the scale factor $a(t)$ of the
FLRW
metric, or equivalently by the Hubble parameter $H(t)\equiv a^{-1}da/dt$.
The contribution of a general dark energy, modified gravity, or dark sector interaction model on this evolution can be characterized in terms of an effective fluid with energy-momentum tensor $T_{\rm eff}^{\mu\nu} \equiv \kappa^{-2}G^{\mu\nu}-T_{\rm m}^{\mu\nu}$, where $\kappa^2=8\pi G$ with bare gravitational constant $G$. $G^{\mu\nu}$ is the Einstein tensor and $T_{\rm m}^{\mu\nu}$ denotes the matter component including baryonic contributions, cold dark matter, and radiation.

At the background level, the effective fluid is fully specified by its energy density $\bar{\rho}(t)=-T_{{\rm eff}\:0}^0$ and pressure $\bar{p}(t)=T_{{\rm eff}\:i}^i$, which
can be parametrized by the equation of state $w(t)=\bar{p}(t)/\bar{\rho}(t)$.
For the cosmological constant $\Lambda$, the two relate as $\bar{p}_{\Lambda}=-\bar{\rho}_{\Lambda}$ with $w=-1$.
For generic dark energy models, however,
the equation of state departs from that value.
For instance, in quintessence models~\cite{Wetterich1988,Ratra:1988}
$-1 < w(t) \leq 1$ and for a choice of $w(t)$ remaining in this regime, one can always
find a scalar field potential that reproduces the desired equation of state.
More exotic dark energy models are required in order to cover the phantom regime $w<-1$.
This can for instance be achieved with an additional coupling of the scalar field to the metric or matter components.
Generally, one finds
\begin{equation}
 \bar{\rho} = \bar{\rho}_0 \exp \left[ 3 \int_a^1 \frac{1+w(a')}{a'} da' \right] \,, \label{eq:rhow}
\end{equation}
which follows from energy conservation, where $\bar{\rho}_0$ denotes the energy density today.

The most common approach to parametrizing deviations from $\Lambda$CDM in the cosmological background is to adopt the Chevallier-Polarski-Linder~\cite{Chevallier2001,Linder2003} (CPL) equation of state
\begin{equation}
 w(t)=w_0+w_a[1-a(t)] \,.\label{eq:w0wa}
\end{equation}
This relation is tested with geometric probes of the expansion history.
It is also directly tested with the growth of structure if restricting to quintessence models.
Any observational sign of a departure from  $w = -1$ or $w_a = 0$ would provide evidence against $\Lambda$CDM.
With the parametrization in Eq.~(\ref{eq:w0wa}), Eq.~(\ref{eq:rhow}) simplifies to
\begin{equation}
 \bar{\rho} = \bar{\rho}_0 a^{-3(1+w_0+w_a)} \exp[3w_a(a-1)] \,.
\end{equation}

More generally, the functional dependence of the dark energy equation of state $w(a)$ can be explored within a principal component analysis (PCA)~\cite{Crittenden2009}.
Often, however, in explorations of the effects of parametrized modifications of gravity a cosmological background is assumed that is instead equivalent to concordance cosmology ($w=-1$).
This approach is usually observationally motivated and adopted to separate tests of the growth of structure from the background evolution.
But it can also be motivated in view of some modified gravity models, for instance, chameleon gravity, where viable models produce $w\approx-1$~\cite{Brax2008,Lombriser2013c,Ceron-Hurtado2016,Battye2017}.

\subsubsection{Genuine cosmic self-acceleration before/after GW170817} \label{sec:cosmicacceleration}

Traditionally, modifications of gravity in the cosmological context have been motivated as alternative to the cosmological constant or dark energy as explanation for the late-time accelerated expansion.
Besides introducing a dark energy potential or cosmological constant, the accelerated expansion can instead arise from the kinetic terms of a dark energy field as for instance in k-essence models~\cite{ArmendarizPicon1999}.
Such a kinetic self-acceleration furthermore emerges in Kinetic Gravity Braiding models~\cite{Deffayet2010} or the cubic Galileon~\cite{Nicolis2008} model, but these can be viewed as imperfect dark energy fluids that, for instance, do not change the propagation of gravitational waves (see Sec.~\ref{sec:gw}).
A modification of this propagation may in contrast be motivated as a requirement for a genuine modification of gravity~\cite{Linder2015}.
We shall therefore briefly explore how a cosmic acceleration can arise in modified gravity that does not rely on kinetic or potential energy contributions of a new field but can genuinely be attributed to an intrinsic property of the modified gravitational interactions.

In Ref.~\refcite{Wang2012}, it was argued that for cosmic acceleration to be genuinely attributed to a modified gravity effect there should be no accelerated expansion in a frame where the modified field equations have been transformed into the Einstein field equations.
Cosmic acceleration should otherwise be ascribed to a dark energy contribution instead.
Note that the background evolution in this frame is described by the standard Friedmann equations for some exotic non-minimally coupled matter sector.
This classification can be applied to effective field theory (Sec.~\ref{sec:eft}) with an effective conformal (or pseudo-conformal) transformation $\Omega(t)$ that connects the two frames defined at the cosmological background~\cite{Lombriser2015c}, absorbing contributions from both conformal and disformal factors at the non-perturbative level.\footnote{The transformed frame was dubbed the Einstein-Friedmann frame in Ref.~\refcite{Lombriser2015c}, which does not coincide with the Einstein frame unless $c_{\rm T}=1$. An equivalent argument can generally be made in the Einstein frame by instead performing the transformations defined in Ref.~\refcite{Gleyzes2015}.}
One then finds that genuine self-acceleration should imply
\begin{equation}
 \frac{d^2\tilde{a}}{d\tilde{t}^2} = \frac{1}{\sqrt{\Omega}} \left[ \left( 1+\frac{1}{2}\frac{d\ln\Omega}{d\ln a} \right) \frac{d^2a}{dt^2} + \frac{aH^2}{2}\frac{d^2\ln\Omega}{d(\ln a)^2}  \right] \leq 0 \,, \label{eq:selfaccEF}
\end{equation}
where tildes indicate quantities in the transformed frame.
The observed late-time acceleration implies $d^2a/dt^2>0$ for $a\gtrsim0.6$ such that $-d\ln\Omega/d\ln a\sim\mathcal{O}(1)$.
It can furthermore be shown that~\cite{Lombriser2015c}
\begin{equation}
 -\frac{d\ln\Omega}{d\ln a} = \frac{d\ln (G_{\rm eff}/c_{\rm T}^2)}{d\ln a} \sim\mathcal{O}(1) \,, \label{eq:selfacceleration}
\end{equation}
where $G_{\rm eff}$ is an effective gravitational coupling and $c_T$ denotes the speed of gravitational waves.
These can generally both evolve in time.
Hence, genuine cosmic self-acceleration is either due to a change in the strength of gravity or in its speed, assuming that the speed of light in vacuum remains constant.

Since cosmic acceleration is an order unity effect on the cosmological background dynamics, if due to modified gravity, one would naturally expect an effect of similar strength on structure formation.
Importantly, however, the modifications in $c_{\rm T}$ and $G_{\rm eff}$ can cancel out in the observed structure.
This can for instance occur in Horndeski gravity~\cite{Horndeski1974} when~\cite{Lombriser2014,Lombriser2015c}
\begin{equation}
 1-c_{\rm T}^2 = \frac{\Delta G_{\rm eff}}{\Delta G_{\rm eff} +\alpha_B G_{\rm eff}}\frac{d\ln G_{\rm eff}}{d\ln a} \label{eq:darkdegeneracy}
\end{equation}
or when this relation holds approximately within the observational uncertainties.
Here, $\Delta G_{\rm eff} \equiv G_{\rm eff}-G$ and $\alpha_B$ describes the braiding effect between the kinetic contributions of the scalar and metric fields.
Specifically, the condition~(\ref{eq:darkdegeneracy}) follows from imposing a standard Poisson equation and the absence of effective anisotropic stress, which will be discussed in more detail in Sec.~\ref{sec:eft} (see Eq.~(\ref{eq:degeneracy}) in particular).
Importantly, a direct measurement of the gravitational wave speed $c_{\rm T}$ breaks the degeneracy implied by Eq.~(\ref{eq:darkdegeneracy})~\cite{Lombriser2015c}.

The recent gravitational wave observation GW170817~\cite{Abbott2017a} with LIGO \& Virgo emitted by a neutron star merger in the NGC 4993 galaxy and the wealth of near-simultaneous electromagnetic counterpart measurements constrains the relative deviation between the speeds of gravity and light at $\mathcal{O}(10^{-15})$~\cite{Abbott2017b}.
With anticipation of this event and bound~\cite{Nishizawa2014,Lombriser2015c},
implications on cosmic self-acceleration from modified gravity from such a measurement were first analyzed in Refs.~\refcite{Lombriser2015c} and~\refcite{Lombriser2016a} (see Sec.~\ref{sec:gw} for further discussions on the speed of gravitational waves).
In particular, $c_{\rm T}\simeq1$ implies that self-acceleration must be due to an evolving $G_{\rm eff}$ from Eq.~(\ref{eq:selfacceleration}) and that the order unity effect of the gravitational modification can no longer hide
in the cosmic structure due to Eq.~(\ref{eq:darkdegeneracy})~\cite{Lombriser2015c}.
Ref.~\refcite{Lombriser2016a} then showed that in this case the minimal modification of gravity required for self-acceleration in the Horndeski framework (Sec.~\ref{sec:linparam}) provides a $3\sigma$ inferior fit to current cosmological data than the cosmological constant.
The tension is mainly attributed to the cross correlations of the integrated Sachs-Wolfe (ISW) effect in the cosmic microwave background (CMB) temperature anisotropies with foreground galaxies and cannot be evaded with screening mechanisms.
Moreover, a rescaled constraint is also applicable to theories where a self-acceleration could emerge from a self-interaction in the dark matter, which would not affect the baryonic components and Solar System tests.

It should, however, be noted that dark energy fields embedded in the Horndeski theory are still viable candidates to explain the late-time acceleration.
Such fields may also naturally be expected to couple to matter but the corresponding modification of gravity cannot be the main driver for cosmic acceleration.
This also removes a preferred scale for the strength of the coupling.
However, a modification of gravity with a Hubble scale deviation from GR may also be a remnant of a mechanism that tunes the cosmological constant to its observed value, which would indirectly govern cosmic acceleration.
Finally, it is also worth noting that the dark degeneracy reappears in more general theories than Horndeski gravity~\cite{Lombriser2014}, which again allows self-acceleration to be hidden in the large-scale structure and evade the tension in galaxy-ISW cross correlations.
The minimal gravitational modification that is necessary in general scalar-tensor theories to produce genuine cosmic self-acceleration~\cite{Lombriser2016a} is, however, expected to leave a $5\sigma$ tension in Standard Sirens tests since the gravitational waves are not affected by the degeneracy~\cite{Lombriser2015c} (Sec.~\ref{sec:gw}).

\subsection{Linear cosmology} \label{sec:linear}

In addition to the cosmological background evolution, modifications of gravity or exotic dark energy models can manifest in the linear fluctuations around that background.
The incurring effects are manifold and 
extensive efforts have been devoted to
establish
general
parametrization frameworks for the linear cosmological perturbation theory of modified gravity and dark energy models.
Different approaches include the growth-index parametrization (Sec.~\ref{sec:growthindex}), the PPF formalisms based on defining closure relations for the system of differential equations determining the linear cosmological perturbations (Sec.~\ref{sec:closure}), the EFT of dark energy and modified gravity (Sec.~\ref{sec:eft}), and a parametrization for the impact on the cosmological propagation of gravitational waves (Sec.~\ref{sec:gw}).
The parametrizations encountered in these different formalisms are in general free time and scale dependent functions. A few different choices that are frequently adopted shall briefly be discussed in Sec.~\ref{sec:linparam}.

\subsubsection{Growth-index parametrization} \label{sec:growthindex}

Besides testing for a departure from the cosmological constant in the equation of state $w(a)$ with geometric probes (Sec.~\ref{sec:w}),
the growth of large-scale structure is a further source of valuable constraints on dark energy and modified gravity.
GR predicts that the growth function of matter density fluctuations $D(a)$ should behave as~\cite{Peebles1980}
\begin{equation}
 f \equiv \frac{d\ln D(a)}{d\ln a} = \Om(a)^{\tilde{\gamma}}\,, \label{eq:growthindex}
\end{equation}
where $\tilde{\gamma}\approx6/11$ and $\Om(a)\equiv \kappa^2\bar{\rho}_{\rm m}/(3H^2)$.
In many modified gravity theories or dark sector interaction models $\tilde{\gamma}$ departs from this value~\cite{Linder2005}.
Measurements of $\tilde{\gamma}$ therefore serve as a \emph{consistency test} of GR, comparing measurements of geometry and growth (also see Refs.~\citen{Ishak2005,Mortonson2008,Ruiz2014}).
While observationally useful, it is often not a very natural parametrization to encompass the wealth of possible modified gravity and dark interaction models, and $\tilde{\gamma}$ should generally be a non-trivial time and scale dependent function.

Moreover, the growth index parametrization is incomplete.
It does not uniquely specify the linear cosmological fluctuations unless additional assumptions are made for how the perturbations relate.
Specifically, while it parametrizes the growth of structure, the weak gravitational lensing caused by that structure can differ between models.
Furthermore, the definition made in Eq.~\eqref{eq:growthindex} does not capture violations of the conservation of comoving curvature (see Eq.~(\ref{eq:energycon}) in Sec.~\ref{sec:closure}) encountered at near-horizon scales once departing from $\Lambda$CDM.
The growth-index parametrization can, however, be completed by connecting it, for instance, to the closure relations discussed in Sec.~\ref{sec:closure} with an additional parametrization for the evolution of the comoving curvature $\zeta$~\cite{Lombriser2011a}.
Alternatively, Eq.~\eqref{eq:growthindex} can be reinterpreted as the velocity-to-density ratio $\mathcal{F}\equiv-k_{\rm H} V_{\rm m}/\Delta_{\rm m}$, which absorbs the evolution of $\zeta$ and reduces to $f$ on subhorizon scales $k_H \gg 1$ (see Eq.~(\ref{eq:energycon}) in Sec.~\ref{sec:closure})~\cite{Lombriser2013a,Lombriser2015b}, where $k_H \equiv k/(aH)$.

\subsubsection{Closure relations for linear perturbations} \label{sec:closure}

Upon revisiting the effective energy-momentum tensor $T_{\rm eff}^{\mu\nu}$ defined in Sec.~\ref{sec:w},
which is associated with the extra terms encountered in a modified Einstein field equation,
one can educe that
because of the Bianchi identities and the energy-momentum conservation of the matter components, it must also hold that $\nabla_{\mu}T_{\rm eff}^{\mu\nu}=0$.
Hence, the usual cosmological perturbation theory can be applied.
This implies four fluctuations each in the metric and the energy-momentum tensor.
Four of those are fixed by the Einstein and conservation equations and another two are fixed by the adoption of a particular gauge.
Hence, one is left with two undetermined relations, which requires the introduction of two closure relations that are specified by the particular modified gravity or dark energy model.
This then closes the system of differential equations determining the cosmological perturbations.
In practice, the two closure relations are designed such that the effective fluid mimics the relations between the metric and matter fluctuations of the model in question.

This is typically done through the introduction of an effective modification of the Poisson equation and a gravitational slip, or effective anisotropic stress~\cite{Uzan2006,Caldwell2007,Zhang2007,Amendola2007},
\begin{eqnarray}
 k_H^2\Psi & = & -\frac{\kappa^2\bar{\rho}_{\rm m}}{2H^2}\mu(a,k)\Deltam \,, \label{eq:mu} \\
 \Phi & = & -\gamma(a,k)\Psi \,, \label{eq:gamma}
\end{eqnarray}
respectively (also see Refs.~\citen{Zhao2009,Daniel2010,Dossett2011}), where the comoving gauge is adopted here for the matter perturbations.
$\Lambda$CDM is recovered for $\mu=\gamma=1$.
Alternatively, a range of combinations of these equations are used to define the two closure relations with a variety of symbols used as notation for the effective modifications (see Tab.~1 in Ref.~\refcite{Daniel2010} for relations between some of them).
It is also worth noting that in modified gravity theories $\mu$ and $\gamma$ are generally only taking on a simple analytic form in the subhorizon limit.
At superhorizon scales ($k_H\ll1$) the evolution of the perturbations needs to be absorbed into the effective modifications, which may instead more naturally be described as an extra summand in the Poisson equation~\cite{Hu2007}.
In this {\bf article} Eqs.~\eqref{eq:mu} and \eqref{eq:gamma} will be adopted for all scales.
The system determining the evolution of the scalar modes is then closed by the energy-momentum conservation equations,
\begin{eqnarray}
 \Deltam' & = & -k_H\Vm - 3\zeta' \,, \label{eq:energycon} \\
 \Vm' & = & -\Vm + k_H\Psi \,, \label{eq:momentumcon}
\end{eqnarray}
where $\zeta\equiv\Phi-\Vm/(kH)$ denotes the comoving curvature (see Ref.~\refcite{Bonvin2018} for how a breaking of the conservation equations can be tested independently of $\mu$ and $\gamma$).

At subhorizon scales $\zeta'$ can be neglected in the energy conservation equation and together with momentum conservation and Eq.~\eqref{eq:mu}, it follows that
\begin{equation}
 \Deltam'' + \left( 2 + \frac{H'}{H} \right) \Deltam' - \frac{3}{2}\Omega_m(a)\mu(a,k)\Deltam = 0 \,, \label{eq:growtheq}
\end{equation}
which determines the growth of structure.
More generally, Eq.~(\ref{eq:growtheq}) can be written as a first-order nonlinear differential equation for $\mathcal{F}$ without adopting a subhorizon approximation~\cite{Lombriser2015b}, which then reduces to an equation for $f$ when $k_H\gg1$ (Sec.~\ref{sec:growthindex}).
The gravitational slip parameter $\gamma$ quantifies
a deviation of the lensing potential
\begin{equation}
\frac{1}{2} (\Phi-\Psi) = - \frac{1}{2} (1+\gamma) \Psi \label{eq:Phim}
\end{equation}
from the Newtonian potential $\Psi$ governing the gravitational dynamics.
This is in analogy to the parametrized post-Newtonian parameter $\gamma$ in Sec.~\ref{sec:PPN}.
The gravitational lensing due to a matter distribution is then determined from the Poisson equation of the potential in Eq.~\eqref{eq:Phim} with the effective modification replaced by the combination $\Sigma \equiv \frac{1}{2}(1+\gamma) \mu$.
The effective modifications $\mu$ and $\gamma$ are generally time and scale dependent, and some
frequently adopted parametrizations shall briefly be inspected in Sec.~\ref{sec:linparam}.

Finally, it is worth noting that at superhorizon scales, the equations of motion also simplify.
More specifically, from diffeomorphism invariance, metric gravitational theories with energy-momentum conservation satisfy the adiabatic fluctuations of a flat universe with comoving curvature conservation $\zeta'=0$ in the limit of $k\rightarrow0$~\cite{Bertschinger2006}.
It then follows from momentum conservation that
\begin{equation}
 \zeta'' - \frac{H''}{H'}\zeta' = \Phi'' - \Psi' - \frac{H''}{H'}\Phi' - \left( \frac{H'}{H} - \frac{H''}{H'} \right) \Psi \rightarrow 0
\end{equation}
such that $\gamma$ and $H$ determine the evolution of the potentials.
This can be used to map modified gravity and dark energy models onto the effective fluid modifications in Eqs.~\eqref{eq:mu} and \eqref{eq:gamma} using the evolution of their perturbations in the super- and subhorizon limits~\cite{Hu2007,Lombriser2013a,Lombriser2015b}.

\subsubsection{Effective field theory of dark energy and modified gravity} \label{sec:eft}

A more systematic approach to covering the range of possible dark energy and modified gravity models than by introducing two free effective functions of time and scale is through the effective field theory of cosmic acceleration (EFT)~\cite{Creminelli2009,Park2010,Gubitosi2012,Bloomfield2012,Gleyzes2013,Tsujikawa2014,Bellini2014,Gleyzes2014,Lagos2017} (also see Refs.~\citen{Battye2012,Hu2014,Zumalacarregui2017}).
The gravitational action is written here in unitary gauge, where the time coordinate absorbs a scalar field perturbation in the metric $g_{\mu\nu}$, and is built from the combination of geometric quantities that are invariant under time-dependent spatial diffeomorphisms.
Those quantities are then assigned free time-dependent coefficients.
To quadratic order, describing the cosmological background evolution and linear perturbations of the model space, one obtains in the low-energy limit~\cite{Gubitosi2012,Bloomfield2012}
\begin{eqnarray}
 S & = & \frac{1}{2\kappa^2} \int d^4x\sqrt{-g}\left\{ \vphantom{R^{(3)}\hat{M}^2m_2^2} \Omega(t) R - 2\Lambda(t) - \Gamma(t) \delg^{00} + M_2^4(t) (\delg^{00})^2 \right. \nonumber\\
 & &  - \Mbar_1^3(t) \delg^{00} \delta K^{\mu}_{\ \mu} - \Mbar_2^2(t) (\delta K^{\mu}_{\ \mu})^2 - \Mbar_2^3(t) \delta K^{\mu}_{\ \nu} \delta K^{\nu}_{\ \mu} \nonumber\\
 & & \left. + \hat{M}^2(t) \delg^{00} \delta R^{(3)} + m_2^2(t)(g^{\mu\nu} + n^{\mu}n^{\nu}) \partial_{\mu} g^{00} \partial_{\nu} g^{00} \right\} + \Sm\left[\psi_{\rm m};g_{\mu\nu}\right] \,, \label{eq:eftaction}
\end{eqnarray}
adopting the notation of Ref.~\refcite{Lombriser2014}.
$R$ and $R^{(3)}$ are the four-dimensional and spatial Ricci scalars, $K_{\mu\nu}$ denotes the extrinsic curvature tensor, $n^{\mu}$ is the normal to constant-time surfaces, and $\delta$ indicates perturbations around the background.
$\Lambda$CDM is recovered for $\Omega=1$, constant $\Lambda$, and all remaining coefficients vanishing.
For quintessence models, $\Omega=1$ with $\Lambda$ and $\Gamma$ describing the scalar field potential and kinetic terms.
$M_2^2\neq0$ is introduced in k-essence and $\Mbar_1\neq0$ in the cubic Galileon and Kinetic Gravity Braiding models.
All the coefficients except for $m_2$ are used to embed Horndeski theories, however, with the restriction that $2\hat{M}^2 = -\Mbar_3^2 = \Mbar_2^2$~\cite{Gleyzes2013,Bloomfield2013}.
The second condition ensures the restriction to second-order spatial derivatives in the equations of motion
with $2\hat{M}^2 \neq \Mbar_2^2$ in beyond-Horndeski theories~\cite{Gleyzes2013}.
Finally, $m_2\neq0$ is introduced in Lorentz covariance violating Ho\v{r}ava-Lifshitz gravity~\cite{Horava2009}.

There is a total of nine coefficients in the action~\eqref{eq:eftaction}, where the scale factor $a(t)$ or the Hubble parameter $H$ of the
spatially homogeneous and isotropic background
adds a tenth function.
For simplicity spatial flatness and a matter-only universe with pressureless dust will be assumed here.
The Friedmann equations,
\begin{equation}
  H^2\left(1 + \frac{\Omega'}{\Omega}\right) = \frac{\kappa^2\rhom + \Lambda + \Gamma}{3\Omega} \,, \quad
  \left(H^2\right)' \left(1 + \frac{1}{2}\frac{\Omega'}{\Omega}\right) + H^2\left(3 + \frac{\Omega''}{\Omega} + 2\frac{\Omega'}{\Omega}\right) = \frac{\Lambda}{\Omega} \,, \quad \label{eq:friedmann}
\end{equation}
following from variation of the action with respect to the metric, then provide two constraints between the first three background coefficients in Eq.~\eqref{eq:eftaction}.
Hence, for specified matter content and spatial curvature, the space of dark energy and modified gravity models embodied by Eq.~\eqref{eq:eftaction} can be characterized by eight free functions of time.
In particular, the cosmological background evolution and linear perturbations of Horndeski theories is described by five coefficients only.

Similarly to Eq.~\eqref{eq:eftaction}, the effective action can also be built from the geometric quantities that can be introduced in an Arnowitt-Deser-Misner (ADM) 3+1 decompostion of spacetime with the uniform scalar field hypersurfaces as the constant time hypersurfaces~\cite{Gleyzes2013,Bellini2014}.
Variation of the action then defines an equivalent set of functions describing the cosmological background as well as the scalar and tensor modes of the perturbations.
The formalism separates out the expansion history $H$ as the free function determining the cosmological background evolution, which relates to $\Omega$, $\Gamma$, and $\Lambda$ through Eqs.~(\ref{eq:friedmann}).
The linear fluctuations of Horndeski scalar-tensor theories are then characterized by four time-dependent functions:
the kineticity $\aK \equiv (\Gamma+4M_2^4)/(H^2\kappa^2M^2)$ that parametrizes the contribution of a kinetic energy of the scalar field;
the evolution rate of the gravitational coupling $\aM \equiv (M^2)'/M^2$ with effective Planck mass $M^2 \equiv \kappa^{-2}(\Omega + \Mbar_2^2)$;
the braiding parameter $\aB \equiv (H\Omega' + \Mbar_1^3)/(2H \kappa^2M^2)$ describing the braiding or mixing of the kinetic contributions of the scalar and metric fields;
and the alteration in the speed of gravity $\aT \equiv -(\Mbar_2^2)/(\kappa^2M^2)$ with $c_{\rm T}^2=1+\alpha_{\rm T}$.
Additional terms are introduced if generalizing the formalism to include further higher-derivative scalar-tensor terms, for instance the beyond-Horndeski function $\alpha_{\rm H}\equiv(2\hat{M}^2-\bar{M}^2_2)/(\kappa^2M^2)$~\cite{Gleyzes2013}, or encompass vector-tensor and tensor-tensor theories~\cite{Lagos2017}.
The formalism has also been extended to encompass more general dark sector interaction models~\cite{Gleyzes2015}.
$\Lambda$CDM is recovered when $\alpha_i=0$ $\forall i$.

The perturbed modified Einstein and scalar field equations as well as a reduced system of differential equations in this formalism can be found in Refs.~\refcite{Bellini2014} or \refcite{Lombriser2015b}.
In particular, in Horndeski gravity ($\alpha_H=0$) at subhorizon scales,
time derivatives of the metric potentials and large-scale velocity flows can be neglected with respect to spatial derivatives and matter density fluctuations.
At leading order in $k$, one obtains for the effective modifications 
\bqa
 \mu_{\rm QS} & = & \frac{2\left[\aB(1+\aT)-\aM+\aT\right]^2 + \alpha(1+\aT) c_{\rm s}^2}{\alpha \cs^2 \kappa^2M^2} \,, \label{eq:muQS} \\
 \gamma_{\rm QS} & = & \frac{2\aB \left[\aB(1+\aT)-\aM+\aT\right] + \alpha c_{\rm s}^2}{2\left[\aB(1+\aT)-\aM+\aT\right]^2 + \alpha (1+\aT) \cs^2} \,,
\eqa
where $\alpha \equiv 6 \aB^2 + \aK$ and the sound speed of the scalar mode is
\bq
 \cs^2 = -\frac{2}{\alpha} \left[ \vphantom{\frac{1'}{1}} \aB' + (1 + \aT)(1 + \aB)^2 - \left(1 + \aM - \frac{H'}{H}\right)(1 + \aB) + \frac{\rhom}{2H^2 M^2} \right] \,. \label{eq:cs}
\eq
One finds that $\alpha_K$, and hence $M_2^4$, does not contribute in the subhorizon limit, but it can give rise to a clustering effect on very large scales~\cite{Lombriser2014}.
Importantly, when $\alpha_H\neq0$, the velocity field and time derivatives contribute at leading order in the subhorizon limit of the field equations such that a quasistatic approximation of $\mu$ and $\gamma$ becomes inaccurate without additional information on the growth rate of matter density fluctuation $f$ or time derivatives of $\Phi$ and $\zeta$~\cite{Lombriser2015b}.

Stability of the background cosmology to the scalar and tensor modes requires~\cite{Bellini2014}
\begin{equation}
 \frac{M^2\alpha}{(1+\alpha_{\rm B})^2} > 0 \,, \quad c_{\rm s}^2>0 \,, \quad M^2>0 \,, \quad c_{\rm T}^2 > 0 \,, \label{eq:stability}
\end{equation}
which also implies $\alpha>0$.

Finally, it should be noted that there are more free functions in the theory space spanned by the coefficients in Eq.~\eqref{eq:eftaction}, or $H$ and $\alpha_i$, than what can be measured by geometric probes and large-scale structure observations.
In particular, this allows for a model space that is degenerate with $\Lambda$CDM~\cite{Lombriser2014}, producing the same expansion history and linear scalar fluctuations as standard cosmology despite $\alpha_i\neq0$, which also contains genuinely self-accelerated models (Sec.~\ref{sec:cosmicacceleration})~\cite{Lombriser2015c}.
For Horndeski theories one of the degeneracy conditions is~\cite{Lombriser2014,Lombriser2015c} (cf.~Eq.~(\ref{eq:darkdegeneracy}) in Sec.~\ref{sec:cosmicacceleration})
\begin{equation}
 \alpha_{\rm T} = \frac{\kappa^2M^2-1}{(1+\alpha_{\rm B})\kappa^2M^2-1}\alpha_{\rm M} \,, \label{eq:degeneracy}
\end{equation}
where self-acceleration is produced at late times for~\cite{Lombriser2015c} (Sec.~\ref{sec:cosmicacceleration})
\begin{equation}
 \frac{\Omega'}{\Omega} = \aM + \frac{\aT'}{1+\aT} \lesssim -\mathcal{O}(1) \,. \label{eq:selfacc}
\end{equation}
As we shall see next, testing the modifications in the propagation of gravitational waves breaks the degeneracy in Eqs.~(\ref{eq:degeneracy}) and (\ref{eq:selfacc}).

\subsubsection{Cosmological propagation of gravitational waves} \label{sec:gw}

With the direct detection of gravitational waves, further constraints on linear cosmological modifications of gravity can be obtained from their effects on the propagation of the wave.
A parametrization for these modifications was introduced in Ref.~\refcite{Saltas2014} with the wave equation
\bq
 h_{ij}'' + \left(3 + \frac{H'}{H} + \nu \right)h_{ij}' + \left(c_{\rm T}^2k_H^2 + \frac{\tilde{\mu}^2}{H^2} \right) h_{ij} = \frac{\Gamma}{H^2} \gamma_{ij} \,, \label{eq:gw}
\eq
where $h_{ij}\equiv g_{ij}/g_{ii}$ is the linear traceless spatial tensor perturbation.
In addition to a possible impact on the expansion history $H$, the gravitational modifications are characterized by the Planck mass evolution rate $\nu$, the speed of the wave $c_{\rm T}$, the mass of the graviton $\tilde{\mu}$, and the source term $\Gamma\gamma_{ij}$.
The modifications can generally be time and scale dependent.
GR is recovered in the limit of $\nu=\tilde{\mu}=\Gamma=0$ and $c_{\rm T}=1$.
The effects of these modifications on the gravitational waveform have been studied in detail in Ref.~\refcite{Nishizawa2017}, where also the connection to parametrizations in the strong-field regime have been considered (see Sec.~\ref{sec:waveforms}).
Vector-tensor theories, for instance, can introduce $c_{\rm T}\neq1$ with $\nu=\tilde{\mu}=\Gamma=0$, and bimetric massive gravity produces $\tilde{\mu}\neq0$, $\Gamma\gamma_{ij}\neq0$, $\nu=0$, and $c_{\rm T}=1$~\cite{Saltas2014}.
Here, we shall focus on scalar-tensor theories, which are described by $\nu=\alpha_{\rm M}$, $c_{\rm T}^2=1+\alpha_{\rm T}$, and $\tilde{\mu}=\Gamma=0$, directly relating the modifications in the propagation of the wave to the modifications in the large-scale structure (Sec.~\ref{sec:eft}) as well as the modifications required for a genuine self-acceleration effect in Eq.~(\ref{eq:selfacc}).

Particular attention has been given to constraints on $c_{\rm T}$.
For instance, the observation of ultra high energy cosmic rays implies a strong constraint on gravitational Cherenkov radiation from a subluminal propagation of the waves as otherwise the radiation would decay at a rate proportional to the square of their energy $\mathcal{O}(10^{11}~\textrm{GeV})$ and not reach Earth~\cite{Moore2001,Caves1980}.
For galactic $\mathcal{O}(10~\textrm{kpc})$ or cosmological $\mathcal{O}(1~\textrm{Gpc})$ origin,
the relative deviation in $c_{\rm T}$
is constrained to be smaller than $\mathcal{O}(10^{-15})$ or $\mathcal{O}(10^{-19})$, respectively.
The constraint is, however, only applicable to subluminal deviations.
Conservatively, the effect could furthermore be screened in case of galactic origin and the constraints inferred from the short wavelengths of the gravitational
waves may generally not be applicable to the low-energy effective modifications considered in the cosmological background and large-scale structure~\cite{Jimenez2015,Lombriser2015c,Battye2018}.
The first caveat is evaded by bounds on the energy loss in pulsars, which constrains both subluminal and superluminal deviations in $c_{\rm T}$ at the subpercent level for Horndeski theories that rely on Vainshtein screening~\cite{Jimenez2015,Brax2015}.
In particular this implies the inviability of Galileon theories~\cite{Lombriser2015c} if combined with deficiencies in the large-scale structure~\cite{Barreira2014b} such as an observational incompatibility of galaxy-ISW cross correlations~\cite{Barreira2014b,Lombriser2009,Kimura2011},
provided the applicability of the cosmological time variation of the scalar field at the binary system.
However, binary pulsars may conservatively still be considered screened by other shielding mechanisms for more general theories.
Both the first and second caveat can be avoided by a direct cosmological measurement of $c_{\rm T}$ over linear, unscreened distance scales.

It was first anticipated in Refs.~\refcite{Nishizawa2014} and \refcite{Lombriser2015c} that a constraint on the relative deviation between the speeds of gravity and light of $\mathcal{O}(10^{-15})$ should be obtained from the comparison of the arrival times between a gravitational wave measured in LIGO \& Virgo and the electromagnetic counterparts emitted by a neutron star merger or a merger between a neutron star and a black hole.
This is a direct consequence of resolution of the detectors limited to $\mathcal{O}(100~\textrm{Mpc})$ for such an event and from higher likelihood of an event at larger volume with larger distances as well as emission time uncertainties of $\mathcal{O}(1~s)$.
It was also estimated that a few such events with simultaneous signals should be expected per year of operation of the detectors.
Such a measurement has recently been realized with GW170817~\cite{Abbott2017a} and the wealth of counterpart
observations~\cite{Abbott2017b}, providing a constraint in agreement with the predictions of Refs.~\refcite{Nishizawa2014} and \refcite{Lombriser2015c}.
In particular, for Horndeski theory the nearly simultaneous arrival implies a breaking of the degeneracy in the large-scale structure with $\alpha_{\rm T}\simeq0$ and that a genuine cosmic self-acceleration effect is incompatible with the observed cosmic structure~\cite{Lombriser2015c,Lombriser2016a} (Secs.~\ref{sec:cosmicacceleration} and \ref{sec:eft}).
Specifically, tensions arise in the observed galaxy-ISW cross correlations that directly test the evolving gravitational coupling $\alpha_{\rm M}$ yielding self-acceleration through its impact on the evolving gravitational potentials.
Furthermore, $c_T=1$ implies $G_{4X}=G_5=0$ in the Horndeski Lagrangian,
first pointed out in the context of the gravito-Cherenkov constraints~\cite{Kimura2011b} and discussed for the arrival time measurements in Ref.~\refcite{McManus2016}.\footnote{Note that while a constant $G_5$ does not modify $c_{\rm T}$, its contribution to the action can be removed by integration by parts and can therefore be considered vanishing. It can furthermore be combined with a constant $G_3$ to form the unique divergence-free tensor that can be contracted with
$\partial^{\mu}\phi\partial^{\nu}\phi$, i.e., the Einstein tensor $G_{\mu\nu}$ from $\mathcal{L}_5$ with the contribution of $\Lambda g_{\mu\nu}$ from $\mathcal{L}_3$,
and that can hence be removed by integration by parts.
}
Finally, it is also worth noting that the GW170817 constraint arises from a merger in
the NGC 4993 galaxy of
the Hydra constellation at $z=0.01$.
While applicable for tests of cosmic self-acceleration, operating in the same redshift regime, the speeds of gravity and light may not necessarily be equal at higher redshifts, also unleashing the possibility of a dark degeneracy in the high-redshift cosmic structure.

Further constraints on $c_{\rm T}$ have been discussed as forecasts for arrival time comparisons with  supernovae emissions~\cite{Nishizawa2014,Saltas2014,Lombriser2015c,Brax2015}, which however are limited to galactic events that may be screened and are limited to a low rate of a few events per century.
LIGO black hole mergers~\cite{Cornish2017} or the comparison between the arrival times of the waves in the detectors~\cite{Blas2016} only provide weak constraints on $c_{\rm T}$.
Forecasts have also been discussed for eclipsing binary systems observable with the LISA detectors~\cite{Bettoni2016}.
Finally, order unity bounds on $c_{\rm T}$ can also be placed at early times from the B-mode power spectrum of the CMB~\cite{Raveri2014}.

Besides the speed of the gravitational waves, further constraints can be inferred for the modified damping term in Eq.~(\ref{eq:gw}) with the running Planck mass adding to the Hubble friction.
For early-time modifications, effects on the CMB B-modes have been employed for constraints in Ref.~\refcite{Amendola2014}.
For bounds on low-redshift modifications relevant to cosmic acceleration (Sec.~\ref{sec:cosmicacceleration}), Ref.~\refcite{Saltas2014} suggested the use of Standard Sirens~\cite{Schutz1986,Holz2005} and first forecasts have been inferred in Refs.~\refcite{Lombriser2015c,Belgacem2017}, and \refcite{Amendola2017}.
More specifically, gravitational waves provide a distance measurement by the decay of the wave amplitude with the luminosity distance.
Combined with an identification of the redshift of the source this therefore provides a luminosity distance-redshift relation $d^{\rm GW}_{\rm L}(z)$, or a \emph{Standard Siren}.
Due to the additional damping of the wave with the running Planck mass on top of the Hubble friction, a comparison of $d^{\rm GW}_{\rm L}(z)$ inferred from Standard Sirens and $d_{\rm L}(z)$ inferred, for instance, from electromagnetic standard candles such as Type Ia supernovae yields a constraint on $\nu$ (or $\alpha_{\rm M}$ and $M$)~\cite{Saltas2014,Lombriser2015c}.
More specifically,  $d_{\rm L}^{\rm GW}(z) = M(0) d_{\rm L}(z) / M(z)$~\cite{Nishizawa2017,Belgacem2017,Amendola2017}.
GW170817 has provided the first Standard Siren~\cite{Abbott2017c}, yielding a constraint of $\sim10\%$ on $H_0$.
It was shown in Ref.~\refcite{Lombriser2015c} that percent-level constraints on the expansion history at low redshifts from Standard Sirens can yield a conclusive $5\sigma$ tension for the minimal modification in scalar-tensor theories necessary for a genuine cosmic self-acceleration (also see Ref.~\refcite{Lombriser2016a}).
This can be achieved either with LISA~\cite{MacLeod2007,Tamanini2016} or
possibly with LIGO, Virgo, and KAGRA before.
Importantly, this constraint will also be applicable to theories beyond Horndeski gravity with $\alpha_{\rm H}\neq0$, where the dark degeneracy is reintroduced in observations of the large-scale structure~\cite{Lombriser2014}, preventing fully exhaustive and general conclusions without the use of gravitational wave measurements.

\subsubsection{Parametrizing linear modifications} \label{sec:linparam}

In Secs.~\ref{sec:closure}--\ref{sec:gw}, we have seen how a small number of effective modifications introduced with generalized frameworks for linear cosmological perturbation theory can enter as free functions of both time and scale in the computations of the formation of large-scale structure and the propagation of gravitational waves.
Here, we shall briefly discuss different choices that are frequently adopted for the time and scale
dependence of
the two closure relations in Sec.~\ref{sec:closure} as well as for the time dependent functions in the effective field theory formalism (Sec.~\ref{sec:eft}).

For the closure functions $\mu(a,k)$ and $\gamma(a,k)$ and equivalent expressions a range of simple expansions in the scale factor $a$ and the wavenumber $k$, smoothly interpolated bins in $a$ (or $z$) and $k$, or PCA tests have been adopted (see Ref.~\refcite{Koyama2016} for a review).
A motivation for the segregation of the scale-dependence in $\mu$ and $\gamma$ was presented in Ref.~\refcite{Silvestri2013} and shall briefly be summarized here.
With adiabatic initial conditions and the weak equivalence principle, Ref.~\refcite{Silvestri2013} points out that the linear gravitational potentials and density fluctuations are directly related to the initial conditions through the transfer functions $T_{\Phi}$, $T_{\Psi}$, $T_{\Delta}$ such that $\mu \sim k_H^2 T_{\Psi}/T_{\Delta}$ and $\gamma = T_{\Phi}/T_{\Psi}$.
For local four-dimensional theories of gravity, where transfer functions depend on $k^2$ only, and restricting to at most second spatial derivatives in the equations of motion, the quasistatic $\mu$ and $\gamma$ must then be described by five functions
of time $p_i(t)$ only with
\begin{equation}
 \mu(k,t) = \frac{1 + p_3(t)k^2}{p_4(t) + p_5(t) k^2} \,, \quad
 \gamma(k,t) = \frac{p_1(t) + p_2(t) k^2}{1 + p_3(t)k^2} \,. \quad
 \label{eq:mugamma}
\end{equation}
Generally, these functions are applicable at leading order as
further, inverse powers of $k$ can appear by accounting for time derivatives and reducing the modified Einstein and field equations to four equations of motion~\cite{Lombriser2015b}.
Note that a similar form to Eq.~(\ref{eq:mugamma}) for $\mu$ and $\frac{1}{2}(1+\gamma)$ was first proposed in Ref.~\refcite{Bertschinger2008} with $p_4$ and the analog of $p_1$ set to unity and power laws of $a(t)$ adopted for the remaining $p_i$.
One can also notice the possibility of a large-scale modification of gravity with a small-scale $\Lambda$CDM limit of $\mu(k\rightarrow\infty)=\gamma(k\rightarrow\infty)=1$, which can for instance be tested on ultra large scales~\cite{Lombriser2013a,Lombriser2014}.
Often, however, the small-scale limit is adopted, where the mass scale involved is assumed to be small, for instance for a gravitational modification driving cosmic acceleration, with the typical observational probes lying well in the subhorizon limit.
Hence, in this case only time dependent modifications of $\mu$ and $\gamma$ are considered but this does, for instance, not capture viable chameleon models.

Focusing on time-dependent modification only, similar parametrizations have been considered for $\mu(a)$ and $\gamma(a)$ and the EFT functions in Sec.~\ref{sec:eft}, for instance, power laws in $a$ or a proportionality of the deviations $p_i$ to $\Omega_{\rm DE}(a)/\Omega_{\rm DE}(a=1)$~\cite{Simpson2012}, where $\Omega_{\rm DE}(a) \equiv \kappa^2\rho_{\rm DE}/(3H^2)$, in which case tests of $\mu$ and $\gamma$ restrict to only two constants, $p_3(a=1)/p_5(a=1)$ and $p_2(a=1)/p_3(a=1)$.

Typically these modifications have been studied with vanishing modifications at early times (however, see, e.g., Refs.~\citen{Lombriser2011a,Raveri2014,Amendola2014}, and \refcite{Lima2016}).
Dropping the motivation of modified gravity as genuine alternative to dark energy, plausibly given the implications of $c_{\rm T}=1$ for cosmic acceleration discussed in Secs.~\ref{sec:cosmicacceleration} and \ref{sec:gw}, one may argue that parametrizations should be generalized to include early-time effects.
It is worth noting, however, that modifications of gravity may still be indirectly related to cosmic acceleration, for instance, as a non-minimally coupled dark energy field or as a remnant of an effect that tunes the cosmological constant, preventing a vacuum catastrophe, in which case effects may still be expected at late times only.
Contrariwise if focusing on aspects of direct cosmic self-acceleration, one may instead integrate out condition~(\ref{eq:selfaccEF}) to find the minimal running of the Planck mass or the speed of gravity required for a genuine self-acceleration from modifying gravity, using Eq.~(\ref{eq:selfacc}).
With $c_{\rm T}=1$, one finds that~\cite{Lombriser2016a} $M_{\rm min}^2 = \kappa^{-2} (a_{\rm acc}/a)^2 e^{C(\chi_{\rm acc}-\chi)}$ for $a\geq a_{\rm acc}$, where $a_{\rm acc} \equiv [\Omega_{\rm m}/(1-\Omega_{\rm m})/2]$ is the scale factor at which the expansion of the late-time universe becomes positively accelerated, $C\equiv2H_0 a_{\rm acc} \sqrt{3(1-\Omega_{\rm m})}$, and $\chi$ denotes the comoving distance.
$M_{\rm min}^2$ is then fully specified by a given expansion history, e.g., matching $\Lambda$CDM, and does not require an additional choice of parametrization for the time dependence of $\alpha_{\rm M}$.
To minimize the impact of the modification on the large-scale structure, it furthermore follows that $\alpha_{\rm B} = \alpha_{\rm M}$ from Eq.~(\ref{eq:muQS})~\cite{Lombriser2016a}.
To quantify the amount of self-acceleration allowed by an observational dataset, one may then introduce the parametrization
$\alpha_{\rm M} = \lambda \: \alpha_{\rm M, min}$
with constant $\lambda$, where $\lambda>1$ allows sufficient gravitational modification for self-acceleration, $\lambda=1$ is the minimal scenario, $\lambda<1$ needs an additional contribution from dark energy or a cosmological constant, $\lambda=0$ corresponds to $\Lambda$CDM or GR, and $\lambda<0$ corresponds to a scenario where the modification of gravity acts to decelerate the expansion and is counteracted by the introduction of dark energy or a cosmological constant.

Finally, it is worth noting that sampling general EFT or $\alpha_i$ coefficients in comparison to observations can be an inefficient process if additionally imposing the stability of the theory with the criteria in Eq.~(\ref{eq:stability}) as not each sample is guaranteed to yield a stable model.
Moreover the contours of the viable parameter space can as a consequence produce edges and leave $\Lambda$CDM in a narrow corner that may only be sparsely sampled and make a statistical interpretation of evidence against or
in favor of an extended theory of gravity more difficult.
Ref.~\refcite{Kennedy2018} therefore argued that the functions to be sampled for the linear perturbations in EFT adopting $\alpha_{\rm T}=0$ should be $M^2, \: c_{\rm s}^2, \: \alpha>0$ in addition to a constant value for $\alpha_{\rm B}$ today or at an initial epoch.
The $\alpha_i$ functions can be reconstructed from these expressions, where it should be noted that $c_{\rm s}^2$ yields a homogeneous linear second-order differential equation for $B$ with $B'/B\equiv(1+\alpha_{\rm B})$ through Eq.~(\ref{eq:cs}) such that a real and unique solution is guaranteed for a real boundary condition on $\alpha_{\rm B}$.

\subsection{Nonlinear cosmology} \label{sec:nonlinear}

The simple and generalized treatment of the cosmological background (Sec.~\ref{sec:background}) and linear perturbations (Sec.~\ref{sec:linear}) for modified gravity and dark energy models enables consistent and efficient computations of the evolution of the background universe, the growth of large-scale structure from the Hubble scale to a few tens of Mpc, and the cosmological propagation of gravitational waves.
However, the linear framework fails at describing structures at increasingly smaller scales, where there is a great wealth of observational data
already available and becoming available with future galaxy surveys like Euclid~\cite{Laureijs2011} or LSST~\cite{Ivezic2008}.
However, cosmological tests are difficult in the nonlinear regime and
tests of gravity are
additionally
complicated in the presence of screening mechanisms.
Nevertheless, a number of frameworks have been developed
to extend the parametrized treatment of the effects of modified gravity to quasilinear scales with perturbation theory and interpolation functions (Sec.~\ref{sec:perturbations}) as well as to perform a parametrization in the deeply nonlinear regime of cosmological structure formation using spherical collapse and halo model approaches (Sec.~\ref{sec:clusters}).

\subsubsection{Weakly nonlinear regime} \label{sec:perturbations}

To describe the matter density fluctuations in a universe governed by a modified theory of gravity, Ref.~\refcite{Hu2007} proposed a phenomenological extension of the linear parametrized post-Friedmann formalism (Sec.~\ref{sec:closure}) for the modeling of the modified nonlinear matter power spectrum by
\begin{equation}
 P(k,z) = \frac{P_{\rm non-GR}(k,z) + c_{\rm nl}\Sigma^2(k,z)P_{\rm GR}}{1+c_{\rm nl}\Sigma^2(k,z)} \,, \label{eq:Pkznl}
\end{equation}
where $P_{\rm GR}(k,z)$ and $P_{\rm non-GR}(k,z)$ denote the nonlinear matter power spectra with background expansion of the modified model adopting either GR or the gravitational modification in the absence of screening.
The weighting function $\Sigma^2(k,z)$ governs the efficiency of the screening, where $c_{\rm nl}$ controls the scale of the effect and could also be time dependent.
The weight
\begin{equation}
 \Sigma^2(k,z) = \left[\frac{k^3}{2\pi^2}P_{\rm lin}(k,z) \right]^n \label{eq:Sigmaweight}
\end{equation}
was adopted in Refs.~\refcite{Hu2007} and \refcite{Koyama2009} with $P_{\rm lin}(k,z)$ denoting the linear power spectrum of the modified gravity model.
The one-loop perturbations of Dvali-Gabadadze-Porrati~\cite{Dvali2000} (DGP) and $f(R)$ gravity~\cite{Buchdahl1970} are well described by $n=1$ and $n=1/3$, respectively, with a $c_{\rm nl}(z)$ that can be fitted to the computations.
While the combination of Eq.~(\ref{eq:Pkznl}) with the weighting function in Eq.~(\ref{eq:Sigmaweight}) recovers $P(k,z)$ at quasilinear scales to good accuracy, increasingly more complicated $\Sigma^2(k,z)$ need to be devised to reproduce power spectra from $N$-body simulations at increasingly nonlinear scales~\cite{Zhao2011}.

Alternatively, the combination of the spherical collapse model and the halo model (see Sec.~\ref{sec:clusters}) with linear perturbation theory and one-loop computations or a simple quasilinear interpolation motivated by $c_{\rm nl}\Sigma^2(k,z)$ provides an accurate description for $P\left(k\lesssim10~h\,{\rm Mpc}^{-1}\right)$ for a range of
scalar-tensor theories~\cite{Brax2013,Lombriser2013c,Lombriser2014a}.
The combination of generalized perturbative computations to one-loop order with a generalized modified spherical collapse model promises to be a good approach to designing a nonlinear extension to the linear parametrized post-Friedmannian framework~\cite{Lombriser2016b}.
The particulars are, however, still being developed.
It is worth noting that for chameleon, symmetron, and dilaton models a generalized parametrization covering the linear and nonlinear scales has been developed in Ref.~\refcite{Brax2012a} (Sec.~\ref{sec:nonlinearparam1}) and applied to the modeling of the nonlinear matter power spectrum~\cite{Brax2013}.

Recently, there has been much progress on the further generalization of the computation of higher-order perturbations for modified gravity and dark energy theories (see, e.g., Refs.~\citen{Bellini2015,Taruya2016,Bose2016,Yamauchi2017,Frusciante2017,Bose2018,Hirano2018}).
While for general modifications of gravity the next-order perturbations introduce new EFT coefficient to the ones discussed in Sec.~\ref{sec:eft}~\cite{Frusciante2017}, it should be noted that for Horndeski theory these terms depend on $G_{4X}$ and $G_5$~\cite{Bellini2015,Yamauchi2017}, which are vanishing due to the $\alpha_{\rm T}\simeq0$ constraint from GW170817 (Sec.~\ref{sec:gw}) and hence do not expand the parameter space at second-order in the quasistatic perturbations.

\subsubsection{Deeply nonlinear regime} \label{sec:clusters}

The cosmological structure formation in the nonlinear regime can be studied with the spherical collapse model, where a dark matter halo is approximated by a spherically symmmetric top-hat overdensity with its evolution described by the nonlinear continuity and Euler equations from an initial condition to the time of its collapse.
For a metric theory of gravity with a pressureless non-relativistic matter fluid, these equations become~\cite{Peebles1980,Schmidt2008,Pace2010}
\begin{eqnarray}
 \dot{\delta} + \frac{1}{a}\mathbf{\nabla}\cdot(1+\delta)\mathbf{v} & = & 0 \,, \\
 \dot{\mathbf{v}} + \frac{1}{a}\left(\mathbf{v}\cdot\mathbf{\nabla}\right)\mathbf{v} + H \mathbf{v} & = & -\frac{1}{a}\mathbf{\nabla}\Psi \,,
\end{eqnarray}
where $\delta\equiv\drhom/\rhomb$, dots indicate derivatives with respect to physical time, and comoving spatial coordinates have been adopted.
Combining these equations, one obtains
\begin{equation}
 \ddot{\delta} + 2H\dot{\delta} - \frac{1}{a^2}\nabla_i\nabla_j(1+\delta)v^iv^j = \frac{1}{a^2}\nabla_i(1+\delta)\nabla^i\Psi \,.
\end{equation}
For a spherical top-hat density with $\mathbf{v}=A(t)\mathbf{r}$ of amplitude $A$ and from the continuity equation, one finds
\begin{equation}
 \frac{1}{a^2}\nabla_i\nabla_jv^iv^j=\frac{4}{3}\frac{\dot{\delta}^2}{(1+\delta)^2} \,.
\end{equation}
This yields the spherical collapse equation
\begin{equation}
 \ddot{\delta} + 2H\dot{\delta} - \frac{4}{3}\frac{\dot{\delta}^2}{(1+\delta)^2} = \frac{1+\delta}{a^2}\mathbf{\nabla}^2\Psi \,, \label{eq:sphcoll}
\end{equation}
from which one infers the evolution of a spherical shell at the edge of the top hat
\begin{equation}
 \frac{\ddot{\zeta}}{\zeta} = H^2 + \dot{H} - \frac{1}{3a^2}\mathbf{\nabla}^2\Psi \,, \label{eq:shellevol}
\end{equation}
where $\zeta(a)$ denotes the physical top-hat radius at $a$ and we have used that mass conservation implies a constant $M=(4\pi/3)\rhomb(1+\delta)\zeta^3$.

The particular choice of metric gravitational theory enters through the background evolution and the Poisson equation for $\Psi$.
One can generally parametrize the modification of the Poisson equation as
\begin{equation}
 \mathbf{\nabla}^2\Psi \equiv \frac{a^2}{2} \left(1+\frac{\Delta\Geff}{G}\right) \kappa^2 \drhom \,, \label{eq:Geff}
\end{equation}
which yields the spherical collapse equation
\begin{equation}
 \frac{\ddot{\zeta}}{\zeta} = H^2 + \dot{H} - \frac{\kappa^2}{6} \left(1+\frac{\Delta\Geff}{G}\right) \drhom \,. \label{eq:MGsphcoll}
\end{equation}
Interpreting the gravitational modifications as an effective fluid with energy-momentum tensor $T^{\mu\nu}_{\rm eff}$ (Sec.~\ref{sec:w})
the first two terms on the right-hand side of Eq.~(\ref{eq:MGsphcoll}) can be rewritten as $H^2 + \dot{H} = -\kappa^2 \left[ \rhomb + ( 1+3w_{\rm eff}) \bar{\rho}_{\rm eff} \right]/6$ using the Friedmann equations.

One can further define the comoving top-hat radius $\RTH$ with $\zeta(a_i)=a_i\RTH$ at an initial scale factor $a_i\ll1$ and $y\equiv \zeta/(a\,\RTH)$, where mass conservation, $\rhomb a^3 \RTH^3 = \rhom \zeta^3$, implies $\rhom/\rhomb=y^{-3}$.
This yields
\begin{equation}
 y'' + \left(2+\frac{H'}{H}\right) y' + \frac{1}{2}\Om(a)\left(1+\frac{\Delta\Geff}{G}\right) \left(y^{-3}-1\right) y = 0 \,, \label{eq:ydiff}
\end{equation}
which is typically solved by setting initial conditions in the matter-dominated regime with $y_i \equiv y(a_i) = 1 - \delta_i/3$ and $y_i' = - \delta_i/3$.

It is worth noting that in modified gravity models, Birkhoff's theorem can be violated, causing a shell crossing and departure of the overdensity from its initial top-hat profile over time~\cite{Borisov2011,Kopp2013}.
Nevertheless, the top hat still provides a good approximation if additionally accounting for the evolution of the surrounding environmental density with an analogous relation to Eq.~(\ref{eq:ydiff}) and its impact on $G_{\rm eff}$~\cite{Li2011c,Li2012,Lombriser2013b,Lombriser2013c}.

A useful quantity to define for the description of nonlinear structure formation is the spherical collapse density $\deltac(z)$, the extrapolation of the initial overdensity $\delta_i$ leading to collapse in Eq.~(\ref{eq:ydiff}) at redshift $z$ with the linear growth factor $D/D_i\equiv\delta_{\rm lin}/\delta_i$.
$D$ is obtained from solving Eq.~(\ref{eq:MGsphcoll}) in the linearized limit that is time dependent only in GR but can be both time and scale dependent in modified gravity models.
To avoid a scale dependence entering through the extrapolation, one may adopt the GR linear growth factor.
Importantly, it is really the initial overdensity $\delta_i$ that is the relevant quantity for structure formation, which allows one to define this effective extrapolation in this computationally more convenient manner as long as the same extrapolation is also adopted for comparable quantities such as the variance of the linear matter power spectrum.
With $\deltac$ one can then model modified cluster properties such as concentration~\cite{Lombriser2012,Lombriser2013c}, halo bias~\cite{Schmidt2008,Lombriser2013c}, cluster profiles~\cite{Lombriser2011b,Lombriser2012,Gronke2015,Mitchell2018}, or the halo mass function~\cite{Schmidt2008,Schmidt2009,Lombriser2010,Li2011c,Lombriser2013b,Lombriser2013c,Barreira2014a,Cataneo2016,Hagstotz2018}, and similar computations can also be performed for modified void properties~\cite{Clampitt2012,Lam2014}.
Those quantities can then be combined in the halo model~\cite{Peacock2000,Seljak2000,Cooray2002} to compose the modified halo model power spectrum~\cite{Schmidt2008,Schmidt2009,Lombriser2013c,Barreira2014a,Achitouv2015} in the deeply nonlinear regime (also see Ref.~\refcite{Li2011a} for a related approach that can be mapped~\cite{Lombriser2013b,Lombriser2013c} and Ref.~\refcite{Lombriser2014a} for a review).
The one-halo term can also be combined with higher-order perturbations to improve accuracy on quasilinear scales (Sec.~\ref{sec:perturbations}).
For higher efficiency in the modeling of cluster properties and the halo model power spectrum, one may also consider the direct parametrization of $\deltac$ instead of using $G_{\rm eff}$ in Eq.~(\ref{eq:Geff}).
Some fitting functions are, for instance, available for $f(R)$ gravity~\cite{Kopp2013,Achitouv2015,Mead2016}.

It should be noted that at deeply nonlinear cosmological scales baryonic effects become important and need to be accounted for in the comparison of model predictions to observations.
Usually, fitting functions are adopted for this that have been matched to observations and hence may conservatively also be used to model the gaseous and stellar components in modified gravity.
With improved physical description~\cite{Fedeli2014,Mead2016}, however, the baryonic effects may themselves be used as test of gravity~\cite{Terukina2013,Sakstein2016} or to discriminate between universal and matter-specific couplings.
Alternatively, statistical techniques can be applied such as a density weighting in the matter power spectrum that break degeneracies between baryonic effects, the variation of cosmological parameters, and modified gravity signatures that may even be unscreened by the statistic~\cite{Lombriser2015a}.

A general parametrization of $G_{\rm eff}$ in the spherical collapse equation~(\ref{eq:ydiff}), embedding the variety of screening mechanisms encountered in modified gravity theories, has been proposed in Ref.~\refcite{Lombriser2016b}.
The parametrization is modularly built from transitions in the effective gravitational coupling of the form
\begin{equation}
 \frac{\Delta \Geff}{G} \sim b \left(\frac{r}{r_0}\right)^a \left\{\left[1+\left(\frac{r_0}{r}\right)^a\right]^{1/b} - 1 \right\} \label{eq:module} \,,
\end{equation}
which has the limits
\begin{equation}
 \frac{\Delta \Geff}{G} \sim \left\{
 \begin{array}{ll}
  b \left( \frac{r}{r_0} \right)^{a(b-1)/b} \,, & {\rm for} \; (b>0) \bigwedge \left[(r \ll r_0, a>0) \bigvee (r \gg r_0, a<0)\right] \,, \\
 -b \left( \frac{r}{r_0} \right)^a \,, & {\rm for} \; (b<0) \bigwedge \left[(r \ll r_0, a>0) \bigvee (r \gg r_0, a<0) \right] \,, \\
  1 \,, &  {\rm for} \; (r \ll r_0, a<0) \bigvee (r \gg r_0, a>0) \,.
 \end{array}
 \right. \label{eq:limits}
\end{equation}
Here, $r_0$ denotes the screening scale, $a$ determines the radial dependence of the gravitational coupling in the screened regime\footnote{Note that this is not the scale factor of the FLRW metric.} together with the interpolation rate $b$ between the screened and unscreened $G_{\rm eff}$.
The form of the transition in Eq.~(\ref{eq:module}) is motivated by the screening profile of the Vainshtein mechanism in DGP gravity, where the expression becomes exact.
While the range of screening and suppression mechanisms in literature can be mapped onto Eq.~(\ref{eq:module})~\cite{Lombriser2016b}, one may alternatively wish to adopt other transition functions instead.

For some models there are multiple transitions in $G_{\rm eff}$ such as the screening regime on small scales and the Yukawa suppression on large scales encountered in chameleon models.
For general modifications of gravity therefore, one may consider the combination~\cite{Lombriser2016b}
\begin{equation}
 \frac{\Geff}{G} = A + \sum_i^{N_0} B_{i} \prod_j^{N_i} b_{ij} \left(\frac{r}{r_{0ij}}\right)^{a_{ij}} \left\{\left[1+\left(\frac{r_{0ij}}{r}\right)^{a_{ij}}\right]^{1/b_{ij}} - 1 \right\} \,, \label{eq:effscr}
\end{equation}
with integers $i,j>0$.
The parameter $A$ describes the relative deviation to the gravitational constant in the fully screened regime, which can be different from unity for modified gravity models, whereas $B_i$ denotes the enhancement in the fully unscreened limit of a particular transition.

The effective modification can be implemented in the spherical collapse model with the replacement $r\rightarrow\zeta=a\,\RTH\,y$ and $y_0\equiv r_0/(a\,\RTH)$ in Eq.~(\ref{eq:effscr}), defining the $G_{\rm eff}$ of Eq.~(\ref{eq:ydiff}), which yields
\begin{equation}
 \frac{\Delta\Geff}{G} = B\,b \left( \frac{\yhal}{y_0} \right)^a \left\{ \left[ 1 + \left( \frac{y_0}{\yhal} \right)^a \right]^{1/b} - 1  \right\} \label{eq:nlpar1}
\end{equation}
for one element in Eq.~(\ref{eq:effscr}).
A single general element $N_0=N_1=1$ can then be modeled by seven parameters $p_{1-7}$ in addition to $p_0=A$, where
\begin{equation}
 a = \frac{p_1}{p_1-1} p_3 \,, \ \ \ \ \
 b = p_1 \,, \ \ \ \ \
 B = p_2 \label{eq:nlpar2}
\end{equation}
and the dimensionless screening scale is given by
\begin{equation}
 y_0 = p_4 a^{p_5} \left(2G\,H_0 M_{\rm vir}\right)^{p_6} \left(\frac{\yenv}{\yhal}\right)^{p_7} \,. \label{eq:nlpar3}
\end{equation}
For $p_7\neq0$ an environmental dependence enters such that $y$ is solved for the collapsing halo $\yhal$ and for its environment $\yenv$.
The parametrization is general enough to allow a mapping of chameleon, symmetron, Vainshtein, and k-mouflage screening as well as Yukawa suppressions or linear shielding with simple analytic expressions for $p_{1-7}$ that are determined by the model parameters (see Ref.~\refcite{Lombriser2016b}).
For instance, $p_2$ is given by a Brans-Dicke coupling, or more generally the unscreened modification that can be matched to the linear $\mu$ of PPF or EFT in Sec.~\ref{sec:closure}.
Furthermore, $p_4$ includes a dependence on cosmological parameters, usually $p_5=-1$, $p_6=0$ for DGP
and $p_7\neq0$
for chameleon models.
Using the scaling method of Ref.~\refcite{McManus2016} one further finds that the radial dependence of $G_{\rm eff}$ in the screened limit is determined by a particular combination of the powers of the second and first derivative terms, $s$ and $t$, and of the derivative-free terms $u$ dominating the scalar field equation in this regime as well as the dimensionality of the matter distribution $v$, $\Delta\Geff/G\sim r^{(3s+2t+u-v)/(s+t+u)}$~\cite{Lombriser2016b}.
Hence, these powers determine $p_3=(3s+2t+u-v)/(s+t+u)$ and can in principle directly be read off from the action of a theory~\cite{McManus2016}.

The parametrization in Eq.~(\ref{eq:effscr}) enables a generalized computation of the spherical collapse density $\deltac$ in modified gravity or dark sector models from which cluster properties and the halo model matter power spectrum may be computed.
It is worth noting that one can for instance use Eq.~(\ref{eq:nlpar3}) to map $\deltac$ from one set of parameters to another, which may be used to build a direct parametrization of $\deltac$.
One may also use Eq.~(\ref{eq:effscr}) in $N$-body simulations, employing techniques such as developed in Refs.~\refcite{Winther2014} or~\refcite{Mead2014}.

\section{Parametrized Post-Newtonian Formalism} \label{sec:PPN}

Model-independent tests of gravity have very successfully been conducted in the low-energy static regime, where a post-Newtonian expansion can be performed and parametrized for generalizations of the gravitational interactions.
Many stringent constraints on departures from GR have been inferred in this limit, particularly from observations in the Solar System.
The formalism is, however, not suitable for cosmology, where the evolution of the background needs to be accounted for, a problem that has been addressed with the frameworks discussed in Sec.~\ref{sec:ppf} (also see Sec.~\ref{sec:further}).
Moreover, the screening mechanisms motivated by cosmological applications also introduce further complications in this low-energy static expansion.
Sec.~\ref{sec:PN} briefly reviews the post-Newtonian series and Sec.~\ref{sec:PPNexpansion} presents its parametrization for generalized tests of gravity.
Sec.~\ref{sec:PPNscreening} discusses how screening mechanisms can be incorporated in the formalism.
Finally, a parametrization of gravitational waveforms inspired by PPN
is briefly discussed in Sec.~\ref{sec:waveforms}.

\subsection{Post-Newtonian expansion} \label{sec:PN}

Slowly evolving weak-field gravitational phenomena are well described by the low-energy static limit of GR, a regime particularly applicable to Solar-System tests of gravity~\cite{Will1993,Will2014}.
The metric of such a system can be expanded in orders\footnote{For clarity the speed of light $c$ is kept stated explicitly here to emphasize the dimensionless counting of orders in the velocity although $c$ has generally been set to unity (Sec.~\ref{sec:intro}).} of $(v/c)^i$
with $h_{\mu\nu}^{(i)}$, neglecting cosmological evolution and assuming an asymptotic Minkowskian limit such that
\begin{equation}
 g_{\mu\nu} = \eta_{\mu\nu} + h_{\mu\nu} = \eta_{\mu\nu} + h_{00}^{(2)} +  h_{ij}^{(2)} + h_{0j}^{(3)} + h_{00}^{(4)} \,. \label{eq:pnexpansion}
\end{equation}
The virial relation determines the order of the Newtonian potential $U\approx\mathcal{O}(2)$.
The matter density for a perfect non-viscous fluid $\rho$ is of the same order, which follows from the Poisson equation.
The pressure in the Solar System is of the order of the gravitational energy $\rho\,U$ and the specific energy density $\Pi$ is of that of $U$.
Hence, for the energy-momentum tensor, one finds to fourth order,
\begin{eqnarray}
 T_{00} & = & \rho [1 + \Pi + v^2 - h^{(2)}_{00}] \,, \\
 T_0^{\ i} & = & - \rho\, v^i \,, \\
 T^{ij} & = & \rho\, v^i v^j + p\, \delta^{ij} \,.
\end{eqnarray}
With the system slowly evolving in time, it holds that $d/dt = \partial_t + \mathbf{v}\cdot\mathbf{\nabla} \approx 0$, which indicates that spatial derivatives are an order lower than time derivatives.
Adopting the standard post-Newtonian gauge with diagonal and isotropic metric, one finds for GR the conditions
\begin{eqnarray}
 h^\mu_{ i,\mu} - \frac{1}{2}h^\mu_{\mu,i} & = & 0 \,, \label{eq:gauge1gr}\\
 h^\mu_{ 0, \mu} - \frac{1}{2}h^\mu_{\mu,0} & = & -\frac{1}{2}h_{00,0} \,, \label{eq:gauge2gr}
\end{eqnarray}
where the relations can change for modified theories of gravity.

The components of the Ricci tensor $R_{\mu\nu}$ are to
fourth order
\begin{equation}
 2R_{00} = -\mathbf{\nabla}^2 h_{00} - \left(h_{jj,00} - 2h_{j0,j0}\right) + h_{00,j}\left(h_{jk,k} - \frac{1}{2}h_{kk,j}\right) - \frac{1}{2}| \mathbf{\nabla} h_{00} | ^2 + h_{jk}h_{00,jk}
\end{equation}
and to
second order
\begin{eqnarray}
 R_{0j} & = & -\frac{1}{2} \left(\mathbf{\nabla}^2 h_{0j} - h_{k0,jk}+h_{kk,0j} - h_{kj,0k} \right) \,, \\
 R_{ij} & = & -\frac{1}{2} \left( \mathbf{\nabla}^2 h_{ij} - h_{00,ij} + h_{kk,ij} - h_{ki,kj} - h_{kj,ki} \right) \,.
\end{eqnarray}
With the Einstein equations
$R_{\mu\nu} = 8 \pi G ( T_{\mu\nu} - \frac{1}{2}g_{\mu\nu} T)$
one finds at second order from $R_{00}$ that
$-\frac{1}{2}\mathbf{\nabla}^2 h^{(2)}_{00} = 4\pi G \rho$,
where the Newtonian potential can be defined as $U \equiv h^{(2)}_{00}/(2G)$.
Employing the gauge condition~\eqref{eq:gauge1gr} and using that $T_{ij} = 0$ at second order, yields $4 \pi G \rho \delta_{ij} = -\frac{1}{2} \mathbf{\nabla}^2 h^{(2)}_{ij}$ and, hence, $h^{(2)}_{ij} = 2 G U \delta_{ij}$.
For $h^{(3)}_{0j}$ one obtains
\begin{equation}
 8 \pi G \rho v_j = -\frac{1}{2}\mathbf{\nabla}^2 h^{(3)}_{0j} - \frac{1}{2}GU_{,0j} \label{eq:pn0jgr}
\end{equation}
with the gauge conditions~\eqref{eq:gauge1gr} and \eqref{eq:gauge2gr}.
Eq.~\eqref{eq:pn0jgr} is solved with the Green's function for the Poisson equation and by defining the post-Newtonian potentials
\begin{eqnarray}
 \label{eq:Vi}
 V_i & \equiv & \int \frac{\rho' v_i'}{|\mathbf{x}-\mathbf{x}'|} d^3x'  \,,\\
 \label{eq:Wi}
 W_i & \equiv & \int \frac{\rho'[\mathbf{v}'\cdot(\mathbf{x}-\mathbf{x}')](x-x')_i}{|\mathbf{x}-\mathbf{x}'|^3}d^3x'
\end{eqnarray}
such that
$h^{(3)}_{0j} = -\frac{7}{2}G V_j - \frac{1}{2}G W_j$.
This follows from employing the identity
\begin{equation}
 \frac{\partial}{\partial t} \int \rho' f(\mathbf{x},\mathbf{x}') d^3 x = \int \rho' \mathbf{v}' \cdot \mathbf{\nabla}' f(\mathbf{x},\mathbf{x}') d^3x \left[ 1 + \mathcal{O}(2)\right] \,,
\end{equation}
obtained from the vanishing of the total time derivative with $\rho'\equiv\rho(\mathbf{x}',t)$ and $v_i' = \partial x'_i/\partial t$.
Furthermore, one has
\begin{equation}
 R_{00}^{(4)} = -\frac{1}{2}\mathbf{\nabla}^2(h_{00}^{(4)} + 2U^2 - 8 \Phi_2) \,,
\end{equation}
with the definition
\begin{equation}
 \Phi_2 \equiv \int \frac{\rho' U'}{|\mathbf{x} - \mathbf{x}'|}d^3x' \,.
\end{equation}
The 00 matter component is
\begin{equation}
 T_{00}-\frac{1}{2}g_{00}T = \rho \left( v^2 - \frac{1}{2}h_{00}^{(2)} + \frac{1}{2}\Pi  + \frac{3}{2}\frac{p}{\rho} \right)
\end{equation}
and one can define new post-Newtonian potentials\cite{Will1993} $\Phi_3$ and $\Phi_4$, employing the Green's function for the Laplacian, such that
\begin{equation}
 h^{(4)}_{00} = -2G^2U^2 +4G\Phi_1 +4G^2\Phi_2 + 2G\Phi_3 + 6G\Phi_4\,.
\end{equation}

\subsection{Parametrizing the post-Newtonian expansion} \label{sec:PPNexpansion}

The coefficients of the potentials found for the expansion of the metric in Eq.~\eqref{eq:pnexpansion} depend on the particular gravitational theory assumed.
Theory-independent tests of gravity can be performed by adopting a parametrized post-Newtonian (PPN) formalism for the expansion of the metric~\cite{Will1971,Will1993},
\begin{eqnarray}
 g_{00} & = & -1 + 2GU - 2 \beta G^2U^2 - 2 \xi G\Phi_W + (2 \gamma +2 + \alpha_3 + \zeta_1 - 2\xi) G \Phi_1  \nonumber \\
 & & + 2(3\gamma - 2 \beta +1 + \zeta_2 + \xi) G^2 \Phi_2 + 2(1+\zeta_2)G\Phi_3 + 2(\gamma +3 \zeta_4 - 2 \xi)G \Phi_4 \nonumber \\
 & & -(\zeta_1 - 2 \xi)G\mathcal{A} - (\alpha_1 - \alpha_2 - \alpha_3)G w^2 U - \alpha_2 w_i w_j G U_{ij} + (2 \alpha_3 - \alpha_1) w^i G V_i \,, \label{eq:ppng00} \\
 g_{0i} & = & - \frac{1}{2}(4 \gamma +3 + \alpha_1 - \alpha_2 + \zeta_1 - 2 \xi)GV_i - \frac{1}{2}(1+\alpha_2 - \zeta_1 + 2 \xi)G W_i \nonumber \\
 & & - \frac{1}{2}(\alpha_1 - 2 \alpha_2)G w_i U - \alpha_2w_jG U_{ij} \,, \\
 g_{ij} & = & (1+2\gamma G U) \delta_{ij} \label{eq:ppngij}
\end{eqnarray}
with the PPN parameters $\gamma$, $\beta$, $\xi$, $\alpha_n$, and $\zeta_n$.
Physically, $\gamma$ parametrizes the amount of curvature caused by a unit rest mass and is the analog to the gravitational slip parameter in Eq.~(\ref{eq:gamma}) on cosmological scales.
The parameter $\beta$ quantifies the amount of nonlinearity in the gravitational superposition, $\xi$ captures preferred location effects, $\zeta_n$ and $\alpha_3$ describe violations in the conservation of energy, momentum, or angular momentum, and the $\alpha_n$ parametrize preferred frame effects.
Here, $w^i$ is the system velocity in a universal rest frame and new potentials have been introduced that appear, for instance, in vector-tensor or bimetric theories of gravity~\cite{Will1993}.
GR is recovered when $\gamma = \beta = 1$ and all other parameters vanish.
The parameter values for a range of other gravitational theories
can be found in Ref.~\refcite{Will2014}, where also a summary of constraints on the PPN parameters is presented.
In particular, a bound of $|\gamma-1| \lsim 10^{-5}$ in the Solar System was inferred with the Cassini mission from the Shapiro time delay of a radio echo passing the Sun and orbital dynamics~\cite{Bertotti2003}.
This can be compared to the cosmological parameter $\gamma$ in Sec.~\ref{sec:closure}, which implies that for larger cosmological effects modifications of gravity should be scale or environment dependent, for instance, due to a screening mechanism, or that the new field does not couple to baryons.

\subsection{Incorporating screening mechanisms} \label{sec:PPNscreening}

Because of the linearization of the field equations in the post-Newtonian expansion, the nonlinear interactions that cause the screening effects are removed, which precludes a simple direct mapping of the screened models to the PPN formalism and the straightforward comparison to observational parameter bounds.
Screening effects can also depend on ambient density, giving rise to both low-energy limits where screening operates and where it does not.
Different approaches to performing a PPN expansion in the presence of screening mechanisms have been examined in Refs.~\refcite{AvilezLopez2015,McManus2017} and \refcite{Zhang2016}.
Ref.~\refcite{Zhang2016} derived expressions for the parameters $\gamma$ and $\beta$ and the effective gravitational coupling for a variety of scalar-tensor theories that screen at large values of the Newtonian potential by allowing an environmentally dependent mass for the scalar field, where the relations were found in terms of Newtonian and Yukawa potentials restricting to static spherically-symmetric systems.
A Lagrange multiplier method~\cite{Gabadadze2012,Padilla2012} was employed in Ref.~\refcite{AvilezLopez2015} to perform the post-Newtonian expansion for the cubic Galileon model with Vainshtein screening to order $(v/c)^2$.
Expansions based on such transformations can, however, be mathematically involved for more complex gravitational theories and the method has not been extended to large-field screening like the chameleon effect.
A unified, more systematic, and efficient method for determining the effective field equations dominating in regimes of derivative or large-field screening or no screening was developed through a scaling approach in Ref.~\refcite{McManus2016}, where the post-Newtonian expansion to order $(v/c)^4$ for scalar-tensor theories with Vainshtein and chameleon screening implementing this method was presented in Ref.~\refcite{McManus2017}.
Further applications to k-mouflage, linear shielding,
or a Yukawa suppression can be found in Ref.~\refcite{Lombriser2016b}.

Screening mechanisms may be incorporated in the PPN formalism
following two different approaches~\cite{McManus2016}: either by an extension of the formalism through introducing new potentials; or by promoting the PPN parameters to functions of time and space.
In the second approach, the parameters in Eqs.~\eqref{eq:ppng00}--\eqref{eq:ppngij} become functions of the coordinates with $G$ being replaced by an \emph{effective gravitational coupling} $G_{\rm eff}$ and the parameter $\xi$ introduced by preferred location effects with the spatially dependent expansion being promoted to a matrix $\xi_{ij}$ to accommodate the screened models~\cite{McManus2017}.
More precisely, for chameleon and cubic Galileon models, Ref.~\refcite{McManus2017} finds
\begin{eqnarray}
 G_{\rm eff} & = & G^{(0)} +  \alpha^{q}\frac{\psi^{(q,p)}}{2U} \,, \label{eq:GeffPPN} \\
 \gamma & = & 1 -  \frac{   2\alpha^{q}  \psi^{(q,p)}  }{ 2G^{(0)}U  + \alpha^{q}  \psi^{(q,p)}} \,, \\
 \beta & = & 1 + \alpha^q \left( \beta_{\rm BD}^{(q)} + \beta_{\rm Scr}^{(q)} \right) \,, \\
 \xi_{ij} & = & \alpha^q(1-\delta_{ij}) \frac{(\epsilon_{jkl} W_k V_l)_{\rm eff}}{\epsilon_{ikl}V_kW_l} \,, \label{eq:xiij}
\end{eqnarray}
where $\alpha$ is the scaling parameter~\cite{McManus2016,McManus2017}, relating the mass scale introduced with the new field to the Planck mass.
The parameters $q$ and $p$ denote the orders of the two simultaneous expansions in $\alpha$ and in $(v/c)$, respectively, with the leading orders $q=-1/2$ and $p=1$ for the cubic Galileon model and $q=1/(1-n)$ and $p=2/(n-1)$ for chameleon models with Jordan-frame potential $\alpha(\phi-\phi_{\rm min})^n$.
Furthermore, $G^{(0)}=G\phi_0^{-1}$, where $\phi_0$ is the background scalar field and $\psi$ is its perturbation, solved by the corresponding perturbation equations.
In particular, we note the recovery of the GR result $\gamma = 1$ in the screened limit $\alpha \to \infty$ for the cubic Galileon and $\alpha \to 0$ for the chameleon model (as well as the GR results $\Geff=G^{(0)}$, $\beta=1$, and $\xi_{ij}=0$).
This is because of $\alpha^q\rightarrow0$ in Eqs.~({\ref{eq:GeffPPN}})--(\ref{eq:xiij}) for the corresponding values of $q$ in the two models.
The parameter value for $\beta$ in Brans-Dicke gravity is represented by $\beta_{\rm BD}$.
Furthermore, $\beta_{\rm Scr}=\beta_{\rm Cubic}$ or $\beta_{\rm Scr}=\beta_{\rm Cham}$ are additional terms that contribute for the cubic Galileon and chameleon models, respectively, where
\begin{eqnarray}
 \beta_{\rm BD}^{(-\frac{1}{2})} & = & \frac{  -\Phi_{\rm BD}^{(-\frac{1}{2},3)} + G^{(-\frac{1}{2})} \dfrac{\delta h^{(0,4)}_{00}}{\delta G^{(0)}}+\gamma^{(-\frac{1}{2})}\dfrac{\delta h^{(0,4)}_{00}}{\delta \gamma^{(0)}}}{2 G^{(0)2} U^2 + 4 G^{(0)2} \Phi_2} \,, \\
 \beta_{\rm Cubic}^{(-\frac{1}{2})} & = & \frac{ \Phi_{\rm Cubic}^{(-\frac{1}{2},3)} }{2G^{(0)2} U^2 + 4 G^{(0)2} \Phi_2} \,, \\
 \beta_{\rm Cham}^{(\frac{1}{1-n})} & = & \frac{-\Phi_{\rm Cham}^{\left(\frac{1}{1-n},\frac{2n}{n-1}\right)}}{2G^{(0)2}U^2 + 4G^{(0)2}\Phi_2} \,. \label{eq:betacham}
\end{eqnarray}
The new potentials introduced in Eqs.~\eqref{eq:xiij}--\eqref{eq:betacham} can be found in Ref.~\refcite{McManus2017}.

Alternatively to this approach, attributing new potentials to the new corrections of the metric and assigning them their own parameter in the spirit of the PPN formalism, Ref.~\refcite{McManus2016} finds for the cubic Galileon,
\begin{eqnarray}
 g_{00} & = & g^{\rm PPN}_{00} + \sigma_1 \left(\psi^{(-\frac{1}{2},1)} + \tilde{\Phi}_1^{(-\frac{1}{2},3)} - 3 \mathcal{A}_\psi^{(-\frac{1}{2},3)} - \mathcal{B}_\psi^{(-\frac{1}{2},3)}  + 6 G^{(0)} \tilde{\Phi}_2^{(-\frac{1}{2},3)} \right) \nonumber \\
 & & + \sigma_2 \psi^{(-\frac{1}{2},3)}   +\sigma_{\rm Cubic} \Phi_{\rm Cubic}^{(-\frac{1}{2},3)} \,, \\
 g_{0i} & = & g^{\rm PPN}_{0i} + \sigma_1 \left(\frac{1}{2}\mathcal{V}_i + \frac{3}{2} \mathcal{W}_i \right) \,, \\
 g_{ij} & = & g^{\rm PPN}_{ij} - \sigma_1 \psi^{(-\frac{1}{2},1)} \,,
\end{eqnarray}
where $g^{\rm PPN}_{\mu\nu}$ is defined by Eqs.~\eqref{eq:ppng00}--\eqref{eq:ppngij}.
Here all potentials arising from $\psi^{(-\frac{1}{2},1)}$ are parametrized by the coefficient $\sigma_1$, those from $\psi^{(-\frac{1}{2},3)}$ by $\sigma_2$, and from $\Phi^{(-\frac{1}{2},3)}_{\rm Cubic}$ by $\sigma_\emph{cubic}$, where for the cubic Galileon $\sigma_1=\sigma_2=\sigma_{\rm Cubic}=\alpha^{-\frac{1}{2}}$, which thus vanish in the screened limit $\alpha\rightarrow\infty$.

\subsection{Parametrizing gravitational waveforms} \label{sec:waveforms}

From the measured change in the orbital period of binary pulsar systems due to the energy loss through gravitational wave emission one can infer an upper bound on the self-acceleration of the center of mass from violation of momentum conservation, which constrains the PPN parameter $\zeta_2$~\cite{Will2014}.
Binary pulsars, however, only constrain the lowest two orders of a parametrized expansion of the gravitational waveform~\cite{Abbott2016}.

In the same spirit as the PPN formalism, a parametrized post-Einsteinian (ppE) framework was introduced in Ref.~\refcite{Yunes2009} to parametrize the effects of departures from GR in the dynamical strong-field regime on the gravitational waveforms from the binary coalescence of compact objects with
\begin{equation}
 h(f) = \left( 1 + \sum_j \bar{\alpha}_j u^j \right) e^{i\sum_k \bar{\beta}_k u^k} h_{\rm GR}(f),
\end{equation}
where $u\equiv(\pi \mathcal{M} f)^{1/3}$ with chirp mass $\mathcal{M}$ and frequency $f$.
The GR waveform is reproduced when the parameters $\bar{\alpha}_j$ and $\bar{\beta}_k$ vanish.
A particular subclass of the ppE framework is the generalized inspiral-merger-ringdown waveform model~\cite{Li2011b} recently used by the strong-field tests of GR by LIGO~\cite{Abbott2016,Abbott2016b}, where parametrizations of departures from GR are restricted to fractional changes in the parameters that determine the gravitational wave phase,
\begin{eqnarray}
 h(f) & = & e^{i\delta\Phi_{\rm gIMR}} h_{\rm GR}(f) \,, \\
 \delta\Phi_{\rm gIMR} & = & \frac{3}{128\eta} \sum_{i=0}^7 \phi_i \delta\chi_i(\pi \tilde{M} f)^{(i-5)/3}
\end{eqnarray}
with total mass $\tilde{M}$, symmetric mass ratio $\eta=m_1m_2/(m_1+m_2)^2$, and $i$-th order post-Newtonian GR phase $\phi_i$.
The connection of the ppE framework and the generalized inspiral-merger-ringdown waveform model with the parametrization of modified gravitational wave propagation in Sec.~\ref{sec:gw} (and therefore with EFT in Sec.~\ref{sec:eft}) is discussed in Ref.~\refcite{Nishizawa2017}.
For instance, one finds that $\sum_j \bar{\alpha}_j u^j = \ln [M(k,z)/M(k,0)]$ for $\nu \equiv d\ln M^2/d\ln a$ and that $\sum_k \bar{\beta}_k u^k$ is determined by an integration over $c_{\rm T}$ and $\tilde{\mu}$ (see Ref.~\refcite{Nishizawa2017}).
Particularly, for only time-dependent modifications, it follows that $\bar{\alpha}_j=\bar{\beta}_k=0$ except for non-vanishing $\bar{\alpha}_0$ and $\bar{\beta}_{\pm3}$.
This implies that the deviation in $c_{\rm T}$ is of fourth post-Newtonian order, $\tilde{\mu}$ of $\mathcal{O}(1)$, and $\nu$ of $\mathcal{O}(0)$.
Furthermore, the modification $\delta\Phi_{\rm gIMR}$ is equivalent to that found for $\sum_k \bar{\beta}_k u^k$ with $\nu$ being irrelevant due to the absence of a modification of amplitude in the parametrization. 

An interesting open
problem is the incorporation of screening effects in the modifications of the gravitational waveform.
One approach was recently proposed in Ref.~\refcite{McManus2017}, based on the post-Newtonian expansion of Refs.~\refcite{Mirshekari2013} and \refcite{Lang2014},
which compute the gravitational waveform for a compact binary system in scalar-tensor gravity to post-Newtonian order $(v/c)^4$.
After casting the Einstein equations into a \emph{relaxed form} together with harmonic gauge conditions, those are solved as a retarded integral over the past null cone, whereby the integral is split into a near-zone and radiation-zone part~\cite{Epstein1975,Will1996,Pati2002}.
To first approximation the effects of modified gravity are screened in the near zone and only enter in the radiation zone and through the boundary conditions~\cite{McManus2017}.

\section{Further parametrizations} \label{sec:further} \label{sec:nonlinearparam1}

Besides the PPF, EFT, PPN, or ppE formalisms, there are a range of further parametrization frameworks that have been developed for astrophysical tests of gravity.
A few of those alternative approaches shall be briefly discussed here.

The PPN formalism neglects the evolution of the cosmic background and assumes asymptotic flatness, hence, it needs to be adapted for cosmological applications.
Such an extension is provided by the parametrized post-Newtonian cosmology~\cite{Sanghai2017} (PPNC) framework that is based on four free functions of time.
Thereby the post-Newtonian expansion is adopted for small regions of space and then patched together to determine the cosmological large scales.
This is accomplished by performing coordinate transformations with local scale factors that are associated with a global scale factor employing an appropriate set of junction conditions.
Four free functions of time $\{\alpha,\gamma,\alpha_c,\gamma_c\}$ are then introduced to describe modifications of the Poisson equation of the gravitational potentials $\Phi$ and $\Psi$ up to $(v/c)^3$.
Hereby, $\gamma$ and $\alpha$ can be linked to the PPF parameters in Sec.~\ref{sec:closure} but also to the PPN parameters in Sec.~\ref{sec:PPNexpansion}.
A number of dark energy, scalar-tensor, and vector-tensor theories can be described by the formalism.
However, the framework so far does not provide a relativistic completion nor a description for Yukawa interactions or screening mechanisms.

A more direct analogy to the post-Newtonian formalism for relativistic cosmology than the frameworks discussed in Sec.~\ref{sec:ppf} is given by the post-Friedmannian formalism of Ref.~\refcite{Milillo2015}, where the expansion of the metric is performed in
inverse powers of the light speed
in a cosmological setting in Poisson gauge (also see Ref.~\refcite{Thomas2015} for an application).
After some field redefinition and applying the linearized Einstein equations, the formalism reproduces linear cosmological perturbation theory, thus providing a link between the post-Newtonian limit on small scales and the large-scale structure.
A parametrization of the gravitational modifications in the formalism has yet to be developed.

An analogous approach to the effective field theory of dark energy and modified gravity (Sec.~\ref{sec:eft}) was conducted for the perturbations of a static spherically symmetric system in Ref.~\refcite{Kase2014}.
Thereby the ADM spacetime decomposition was employed for a $2+1+1$ canonical formalism that separates out the time and radial coordinates.
The effective action was then built from scalar quantities of the canonical variables of this spacetime foliation and the lapse, which encompasses scalar-tensor theories of gravity.
Ref.~\refcite{Kase2014} derived three background equations that can be used for the generalized study of screening mechanisms as well as the linear perturbations around this background for stability analyses.

Finally, a unified parametrization covering chameleon, dilaton, and symmetron models on large and small cosmological scales as well as in Solar-System and laboratory tests
was introduced in Ref.~\refcite{Brax2012a}.
It uses the cosmological time variation of the mass $m(t)$ of the scalar field $\phi$ and its coupling $\beta(t)$ at the minimum of its effective potential $V_{\rm eff}(\phi)$.
The starting point is the scalar-tensor action in Einstein frame with $\mathcal{L}_{\phi} = -(\nabla\phi)^2/2 - V(\phi)$
and a conformal factor $A(\phi)$ defined by $g_{\mu\nu} = A^2(\phi)\tilde{g}_{\mu\nu}$, where $\tilde{g}_{\mu\nu}$ denotes the Einstein frame metric.
The scalar-field equation in the presence of pressureless matter is
$\Box \phi = \beta\rho_m + dV/d\phi$
with
\begin{equation}
 \beta(\phi) \equiv M_{\rm Pl} \frac{d\ln A}{d\phi} \,,
\end{equation}
where $M_{\rm Pl}$ denotes the bare Planck mass.
This defines an effective scalar field potential
$V_{\rm eff} = V(\phi) + [A(\phi)-1]\rho_m$ with
\begin{equation}
 m^2(\phi) = \left.\frac{d^2V_{\rm eff}}{d\phi^2}\right|_{\phi_{\rm min}} \,,
\end{equation}
where $\nabla^2\phi=V_{{\rm eff},\phi}$ and $\phi_{\rm min}$ is the minimum of the potential $V_{\rm eff}$.
Appropriate choices of $m(\phi)$ and $\beta(\phi)$ recover the chameleon, symmetron, and dilaton models.
Screening occurs in the regime
$|\phi_{\infty}-\phi_c|\ll 2\beta(\phi_{\infty})M_{\rm Pl} \Psi$,
where $\phi_c$ and $\phi_{\infty}$ are the scalar field values in the minima inside and outside of a body.
For $m^2 \gg H^2$, the scalar field also remains at the minimum in the cosmological background such that its cosmological evolution is given by
$d\phi/dt = 3H\beta A \bar{\rho}_m / (m^2 M_{\rm Pl})$.

This allows one to reconstruct the scalar-tensor Lagrangian from $m(t)$ and $\beta(t)$.
Using
\begin{equation}
 \phi(a) - \phi_{\rm i} = \frac{3}{M_{\rm Pl}} \int_{a_{\rm i}}^a da \frac{\beta(a)\bar{\rho}_m(a)}{a\:m^2(a)} \,,
\end{equation}
where $\phi_{\rm i}$ is the initial scalar field value at $a_{\rm i}$, one finds
\begin{eqnarray}
 \int_{\phi_{\rm i}}^{\phi} \frac{d\phi}{\beta(\phi)} = \frac{3}{M_{\rm Pl}^2} \int_{a_{\rm i}}^a da \frac{\bar{\rho}_m(a)}{a\:m^2(a)} \,, \\
 V = V_{\rm ini} - \frac{3}{M_{\rm Pl}^2} \int_{a_{\rm i}}^a da \frac{\beta^2(a)\bar{\rho}_m^2(a)}{a\:m^2(a)} \,.
\end{eqnarray}
Spherical collapse, halo model, and cosmological $N$-body simulations in this unified approach have been studied in Refs.~\refcite{Brax2012b} and \refcite{Brax2013}.
While the approach covers screening effects that operate at large $\Psi$, it does not cover k-mouflage or Vainshtein screening operating through derivatives of $\Psi$.

\section{Summary \& Outlook} \label{sec:discussion}

Intensive efforts have been devoted to the development of generalized frameworks that enable the systematic exploration and testing of the astronomical and cosmological implications of the wealth of proposed modified gravity, dark energy, and dark sector interaction models.
This {\bf article} reviewed a number of different formalisms devised for this purpose.
In the low-energy static limit of gravity, the PPN expansion has been highly successful in conducting model-independent tests of gravity and inferring stringent constraints on departures from GR, mainly from observations in the Solar System.
An overview was presented of the post-Newtonian expansion, its parametrization, and extensions developed to incorporate the screening mechanisms motivated by cosmological modifications of gravity.

In cosmological settings the PPN formalism is not suitable as the expansion of the background needs to be accounted for.
With the objective of imitating the success of PPN, a number of cosmological counterpart frameworks have been developed, generally commonly referred to as PPF formalisms.
These aim at consistently unifying the effects of modified gravity, dark energy, and dark sector interactions on the cosmological background evolution and the formation of linear and nonlinear large-scale structure.
This includes parametrizations of the equation of state, the growth rate of structure, closure relations for cosmological perturbation theory, EFT, generalized perturbation theory, interpolation functions between linear and nonlinear regimes as well as phenomenologically parametrized screening mechanisms.

The new era of gravitational wave astronomy facilitates further powerful tests of gravity.
Modified gravity and novel interactions can manifest both through cosmological propagation effects or in the emitted gravitational waveforms.
A parametrization formalism for the propagation effects has been reviewed and formalisms inspired by PPN for the parametrization of modified waveforms has also briefly been discussed.
It has also been described how these formalisms can be connected to EFT and PPF.

The PPN and PPF formalisms are generally separated frameworks.
It, however, seems feasible to undertake more general steps towards a \emph{unified parametrization for gravity and dark entities} suitable to all scales and types of observations.
More specifically, if restricting to scalar-tensor theories of the chameleon, dilaton, or symmetron types a unification can be realized through a parametrization of the cosmological time variation of the mass of the scalar field and the coupling at the minimum of its effective potential~\cite{Winther2014}.
This reconstructs a Lagrangian that can be used to connect the different parametrization frameworks as it encapsulates the full nonlinear freedom of these models, enabling applications to cosmology, the low-energy static limit as well as gravitational waves and laboratory tests.
More generally, focusing on a new scalar degree of freedom a unification can in principle be achieved~\cite{Lombriser2016b} by performing a reconstruction~\cite{Kennedy2017,Kennedy2018} of general scalar-tensor theories from the linear EFT functions, which can then be connected to nonlinear PPF frameworks as well as the PPN and waveform formalisms.
Further approaches to unifying PPN and PPF such as with PPNC or a post-Friedmannian expansion have also briefly been inspected.
Completing the development of a global framework for tests of gravity and the dark sector is subject to current research.
Applications of such a framework promise to remain a very interesting and active field of research over the next decades with the wealth of high-quality observational data becoming available for tests of gravity, spanning a wide range of scales and encompassing a great variety of observational probes.

\section*{Acknowledgments}

This work was supported by a Swiss National Science Foundation (SNSF) Professorship grant (No.~170547), a SNSF Advanced Postdoc.Mobility Fellowship (No.~161058), and the Science and Technology Facilities Council Consolidated Grant for Astronomy and Astrophysics at the University of Edinburgh.

\bibliographystyle{ws-ijmpd.bst}
\bibliography{parametrizations.bib}

\begin{thebibliography}{100}

\bibitem{Copeland:2006wr}
E.~J. Copeland, M.~Sami and S.~Tsujikawa, {\em Int. J. Mod. Phys.} {\bf D15}
  (2006) 1753, \href{http://arxiv.org/abs/hep-th/0603057}{{\ttfamily
  arXiv:hep-th/0603057 [hep-th]}}.

\bibitem{Clifton2011}
T.~Clifton, P.~G. Ferreira, A.~Padilla and C.~Skordis, {\em Phys. Rept.} {\bf
  513}  (2012) 1, \href{http://arxiv.org/abs/1106.2476}{{\ttfamily
  arXiv:1106.2476 [astro-ph.CO]}}.

\bibitem{Joyce2014}
A.~Joyce, B.~Jain, J.~Khoury and M.~Trodden, {\em Phys. Rept.} {\bf 568}
  (2015) 1, \href{http://arxiv.org/abs/1407.0059}{{\ttfamily arXiv:1407.0059
  [astro-ph.CO]}}.

\bibitem{Koyama2016}
K.~{Koyama}, {\em Reports on Progress in Physics} {\bf 79} (April 2016)
  046902, \href{http://arxiv.org/abs/1504.04623}{{\ttfamily arXiv:1504.04623}}.

\bibitem{Bull2016}
P.~{Bull}, Y.~{Akrami}, J.~{Adamek}, T.~{Baker}, E.~{Bellini}, J.~{Beltr{\'a}n
  Jim{\'e}nez}, E.~{Bentivegna}, S.~{Camera}, S.~{Clesse}, J.~H. {Davis},
  E.~{Di Dio}, J.~{Enander}, A.~{Heavens}, L.~{Heisenberg}, B.~{Hu},
  C.~{Llinares}, R.~{Maartens}, E.~{M{\"o}rtsell}, S.~{Nadathur}, J.~{Noller},
  R.~{Pasechnik}, M.~S. {Pawlowski}, T.~S. {Pereira}, M.~{Quartin},
  A.~{Ricciardone}, S.~{Riemer-S{\o}rensen}, M.~{Rinaldi}, J.~{Sakstein}, I.~D.
  {Saltas}, V.~{Salzano}, I.~{Sawicki}, A.~R. {Solomon}, D.~{Spolyar}, G.~D.
  {Starkman}, D.~{Steer}, I.~{Tereno}, L.~{Verde}, F.~{Villaescusa-Navarro},
  M.~{von Strauss} and H.~A. {Winther}, {\em Physics of the Dark Universe} {\bf
  12} (June 2016) 56, \href{http://arxiv.org/abs/1512.05356}{{\ttfamily
  arXiv:1512.05356}}.

\bibitem{Joyce2016}
A.~Joyce, L.~Lombriser and F.~Schmidt, {\em Ann. Rev. Nucl. Part. Sci.} {\bf
  66}  (2016) 95, \href{http://arxiv.org/abs/1601.06133}{{\ttfamily
  arXiv:1601.06133 [astro-ph.CO]}}.

\bibitem{Ishak2018}
M.~{Ishak}, {\em ArXiv e-prints}  (June 2018)
  \href{http://arxiv.org/abs/1806.10122}{{\ttfamily arXiv:1806.10122}}.

\bibitem{Will1971}
C.~M. {Will}, {\em The Astrophysical Journal} {\bf 163} (February 1971)   611.

\bibitem{Will1993}
C.~M. {Will}, {\em {Theory and Experiment in Gravitational Physics}} March
  1993.

\bibitem{Will2014}
C.~M. {Will}, {\em Living Reviews in Relativity} {\bf 17} (June 2014)  ~4,
  \href{http://arxiv.org/abs/1403.7377}{{\ttfamily arXiv:1403.7377 [gr-qc]}}.

\bibitem{AvilezLopez2015}
A.~{Avilez-Lopez}, A.~{Padilla}, P.~M. {Saffin} and C.~{Skordis}, {\em Journal
  of Cosmology and Astroparticle Physics} {\bf 6} (June 2015)   044,
  \href{http://arxiv.org/abs/1501.01985}{{\ttfamily arXiv:1501.01985 [gr-qc]}}.

\bibitem{McManus2017}
R.~{McManus}, L.~{Lombriser} and J.~{Pe{\~n}arrubia}, {\em Journal of Cosmology
  and Astroparticle Physics} {\bf 12} (December 2017)   031,
  \href{http://arxiv.org/abs/1705.05324}{{\ttfamily arXiv:1705.05324 [gr-qc]}}.

\bibitem{Uzan2006}
J.-P. {Uzan}, {\em ArXiv Astrophysics e-prints}  (May 2006)
  \href{http://arxiv.org/abs/astro-ph/0605313}{{\ttfamily astro-ph/0605313}}.

\bibitem{Caldwell2007}
R.~{Caldwell}, A.~{Cooray} and A.~{Melchiorri}, {\em Physical Review D} {\bf
  76} (July 2007)   023507,
  \href{http://arxiv.org/abs/astro-ph/0703375}{{\ttfamily astro-ph/0703375}}.

\bibitem{Zhang2007}
P.~{Zhang}, M.~{Liguori}, R.~{Bean} and S.~{Dodelson}, {\em Physical Review
  Letters} {\bf 99} (October 2007)   141302,
  \href{http://arxiv.org/abs/0704.1932}{{\ttfamily arXiv:0704.1932}}.

\bibitem{Amendola2007}
L.~{Amendola}, M.~{Kunz} and D.~{Sapone}, {\em Journal of Cosmology and
  Astroparticle Physics} {\bf 4} (April 2008)   013,
  \href{http://arxiv.org/abs/0704.2421}{{\ttfamily arXiv:0704.2421}}.

\bibitem{Hu2007}
W.~{Hu} and I.~{Sawicki}, {\em Physical Review D} {\bf 76} (November 2007)
  104043, \href{http://arxiv.org/abs/0708.1190}{{\ttfamily arXiv:0708.1190}}.

\bibitem{Tegmark2001}
M.~Tegmark, {\em Phys. Rev.} {\bf D66}  (2002)   103507,
  \href{http://arxiv.org/abs/astro-ph/0101354}{{\ttfamily
  arXiv:astro-ph/0101354 [astro-ph]}}.

\bibitem{Baker2012}
T.~{Baker}, P.~G. {Ferreira} and C.~{Skordis}, {\em Physical Review D} {\bf 87}
  (January 2013)   024015, \href{http://arxiv.org/abs/1209.2117}{{\ttfamily
  arXiv:1209.2117 [astro-ph.CO]}}.

\bibitem{Peebles1980}
P.~J.~E. {Peebles}, {\em {The large-scale structure of the universe}} 1980.

\bibitem{Linder2005}
E.~V. {Linder}, {\em Physical Review D} {\bf 72} (August 2005)   043529,
  \href{http://arxiv.org/abs/astro-ph/0507263}{{\ttfamily astro-ph/0507263}}.

\bibitem{Creminelli2009}
P.~{Creminelli}, G.~{D'Amico}, J.~{Nore{\~n}a} and F.~{Vernizzi}, {\em Journal
  of Cosmology and Astroparticle Physics} {\bf 2} (February 2009)   018,
  \href{http://arxiv.org/abs/0811.0827}{{\ttfamily arXiv:0811.0827}}.

\bibitem{Park2010}
M.~{Park}, K.~M. {Zurek} and S.~{Watson}, {\em Physical Review D} {\bf 81}
  (June 2010)   124008, \href{http://arxiv.org/abs/1003.1722}{{\ttfamily
  arXiv:1003.1722 [hep-th]}}.

\bibitem{Gubitosi2012}
G.~{Gubitosi}, F.~{Piazza} and F.~{Vernizzi}, {\em Journal of Cosmology and
  Astroparticle Physics} {\bf 2} (February 2013)   032,
  \href{http://arxiv.org/abs/1210.0201}{{\ttfamily arXiv:1210.0201 [hep-th]}}.

\bibitem{Bloomfield2012}
J.~{Bloomfield}, {\'E}.~{\'E}. {Flanagan}, M.~{Park} and S.~{Watson}, {\em
  Journal of Cosmology and Astroparticle Physics} {\bf 8} (August 2013)   010,
  \href{http://arxiv.org/abs/1211.7054}{{\ttfamily arXiv:1211.7054
  [astro-ph.CO]}}.

\bibitem{Gleyzes2013}
J.~{Gleyzes}, D.~{Langlois}, F.~{Piazza} and F.~{Vernizzi}, {\em Journal of
  Cosmology and Astroparticle Physics} {\bf 8} (August 2013)   025,
  \href{http://arxiv.org/abs/1304.4840}{{\ttfamily arXiv:1304.4840 [hep-th]}}.

\bibitem{Tsujikawa2014}
S.~Tsujikawa, {\em Lect. Notes Phys.} {\bf 892}  (2015) 97,
  \href{http://arxiv.org/abs/1404.2684}{{\ttfamily arXiv:1404.2684 [gr-qc]}}.

\bibitem{Bellini2014}
E.~Bellini and I.~Sawicki, {\em JCAP} {\bf 1407}  (2014)   050,
  \href{http://arxiv.org/abs/1404.3713}{{\ttfamily arXiv:1404.3713
  [astro-ph.CO]}}.

\bibitem{Gleyzes2014}
J.~Gleyzes, D.~Langlois and F.~Vernizzi, {\em Int. J. Mod. Phys.} {\bf D23}
  (2015)   1443010, \href{http://arxiv.org/abs/1411.3712}{{\ttfamily
  arXiv:1411.3712 [hep-th]}}.

\bibitem{Lagos2017}
M.~{Lagos}, E.~{Bellini}, J.~{Noller}, P.~G. {Ferreira} and T.~{Baker}, {\em
  ArXiv e-prints}  (November 2017)
  \href{http://arxiv.org/abs/1711.09893}{{\ttfamily arXiv:1711.09893 [gr-qc]}}.

\bibitem{Koyama2009}
K.~{Koyama}, A.~{Taruya} and T.~{Hiramatsu}, {\em Physical Review D} {\bf 79}
  (June 2009)   123512, \href{http://arxiv.org/abs/0902.0618}{{\ttfamily
  arXiv:0902.0618 [astro-ph.CO]}}.

\bibitem{Brax2012a}
P.~{Brax}, A.-C. {Davis}, B.~{Li} and H.~A. {Winther}, {\em Physical Review D}
  {\bf 86} (August 2012)   044015,
  \href{http://arxiv.org/abs/1203.4812}{{\ttfamily arXiv:1203.4812
  [astro-ph.CO]}}.

\bibitem{Bellini2015}
E.~{Bellini}, R.~{Jimenez} and L.~{Verde}, {\em Journal of Cosmology and
  Astroparticle Physics} {\bf 5} (May 2015)   057,
  \href{http://arxiv.org/abs/1504.04341}{{\ttfamily arXiv:1504.04341}}.

\bibitem{Taruya2016}
A.~{Taruya}, {\em Physical Review D} {\bf 94} (July 2016)   023504,
  \href{http://arxiv.org/abs/1606.02168}{{\ttfamily arXiv:1606.02168}}.

\bibitem{Bose2016}
B.~{Bose} and K.~{Koyama}, {\em Journal of Cosmology and Astroparticle Physics}
  {\bf 8} (August 2016)   032,
  \href{http://arxiv.org/abs/1606.02520}{{\ttfamily arXiv:1606.02520}}.

\bibitem{Yamauchi2017}
D.~{Yamauchi}, S.~{Yokoyama} and H.~{Tashiro}, {\em Physical Review D} {\bf 96}
  (December 2017)   123516, \href{http://arxiv.org/abs/1709.03243}{{\ttfamily
  arXiv:1709.03243}}.

\bibitem{Frusciante2017}
N.~{Frusciante} and G.~{Papadomanolakis}, {\em Journal of Cosmology and
  Astroparticle Physics} {\bf 12} (December 2017)   014,
  \href{http://arxiv.org/abs/1706.02719}{{\ttfamily arXiv:1706.02719 [gr-qc]}}.

\bibitem{Bose2018}
B.~{Bose}, K.~{Koyama}, M.~{Lewandowski}, F.~{Vernizzi} and H.~A. {Winther},
  {\em ArXiv e-prints}  (February 2018)
  \href{http://arxiv.org/abs/1802.01566}{{\ttfamily arXiv:1802.01566}}.

\bibitem{Hirano2018}
S.~{Hirano}, T.~{Kobayashi}, H.~{Tashiro} and S.~{Yokoyama}, {\em ArXiv
  e-prints}  (January 2018) \href{http://arxiv.org/abs/1801.07885}{{\ttfamily
  arXiv:1801.07885}}.

\bibitem{Zhao2011}
G.-B. {Zhao}, B.~{Li} and K.~{Koyama}, {\em Physical Review D} {\bf 83}
  (February 2011)   044007, \href{http://arxiv.org/abs/1011.1257}{{\ttfamily
  arXiv:1011.1257 [astro-ph.CO]}}.

\bibitem{Lombriser2016b}
L.~{Lombriser}, {\em Journal of Cosmology and Astroparticle Physics} {\bf 11}
  (November 2016)   039, \href{http://arxiv.org/abs/1608.00522}{{\ttfamily
  arXiv:1608.00522}}.

\bibitem{Hu2017}
B.~Hu, X.-W. Liu and R.-G. Cai, {\em Mon. Not. Roy. Astron. Soc.} {\bf 476}
  (2018) L65, \href{http://arxiv.org/abs/1712.09017}{{\ttfamily
  arXiv:1712.09017 [astro-ph.CO]}}.

\bibitem{Sanghai2017}
V.~A.~A. {Sanghai} and T.~{Clifton}, {\em Classical and Quantum Gravity} {\bf
  34} (March 2017)   065003, \href{http://arxiv.org/abs/1610.08039}{{\ttfamily
  arXiv:1610.08039 [gr-qc]}}.

\bibitem{Milillo2015}
I.~{Milillo}, D.~{Bertacca}, M.~{Bruni} and A.~{Maselli}, {\em Physical Review
  D} {\bf 92} (July 2015)   023519,
  \href{http://arxiv.org/abs/1502.02985}{{\ttfamily arXiv:1502.02985 [gr-qc]}}.

\bibitem{Abbott2016a}
B.~P. {Abbott}, R.~{Abbott}, T.~D. {Abbott}, M.~R. {Abernathy}, F.~{Acernese},
  K.~{Ackley}, C.~{Adams}, T.~{Adams}, P.~{Addesso}, R.~X. {Adhikari} and
  et~al., {\em Physical Review Letters} {\bf 116} (February 2016)   061102,
  \href{http://arxiv.org/abs/1602.03837}{{\ttfamily arXiv:1602.03837 [gr-qc]}}.

\bibitem{Abbott2017}
 Virgo, LIGO Scientific Collaboration (B.~P. Abbott {\em et~al.}), {\em Phys.
  Rev. Lett.} {\bf 119}  (2017)   141101,
  \href{http://arxiv.org/abs/1709.09660}{{\ttfamily arXiv:1709.09660 [gr-qc]}}.

\bibitem{Yunes2009}
N.~{Yunes} and F.~{Pretorius}, {\em Physical Review D} {\bf 80} (December 2009)
    122003, \href{http://arxiv.org/abs/0909.3328}{{\ttfamily arXiv:0909.3328
  [gr-qc]}}.

\bibitem{Li2011b}
T.~G.~F. {Li}, W.~{Del Pozzo}, S.~{Vitale}, C.~{Van Den Broeck}, M.~{Agathos},
  J.~{Veitch}, K.~{Grover}, T.~{Sidery}, R.~{Sturani} and A.~{Vecchio}, {\em
  Physical Review D} {\bf 85} (April 2012)   082003,
  \href{http://arxiv.org/abs/1110.0530}{{\ttfamily arXiv:1110.0530 [gr-qc]}}.

\bibitem{Mirshekari2013}
S.~{Mirshekari} and C.~M. {Will}, {\em Physical Review D} {\bf 87} (April 2013)
    084070, \href{http://arxiv.org/abs/1301.4680}{{\ttfamily arXiv:1301.4680
  [gr-qc]}}.

\bibitem{Lang2014}
R.~N. {Lang}, {\em Physical Review D} {\bf 89} (April 2014)   084014,
  \href{http://arxiv.org/abs/1310.3320}{{\ttfamily arXiv:1310.3320 [gr-qc]}}.

\bibitem{Abbott2016}
B.~P. {Abbott}, R.~{Abbott}, T.~D. {Abbott}, M.~R. {Abernathy}, F.~{Acernese},
  K.~{Ackley}, C.~{Adams}, T.~{Adams}, P.~{Addesso}, R.~X. {Adhikari} and
  et~al., {\em Physical Review Letters} {\bf 116} (June 2016)   221101,
  \href{http://arxiv.org/abs/1602.03841}{{\ttfamily arXiv:1602.03841 [gr-qc]}}.

\bibitem{Abbott2016b}
 Virgo, LIGO Scientific Collaboration (B.~P. Abbott {\em et~al.}), {\em Phys.
  Rev.} {\bf X6}  (2016)   041015,
  \href{http://arxiv.org/abs/1606.04856}{{\ttfamily arXiv:1606.04856 [gr-qc]}}.

\bibitem{Saltas2014}
I.~D. {Saltas}, I.~{Sawicki}, L.~{Amendola} and M.~{Kunz}, {\em Physical Review
  Letters} {\bf 113} (November 2014)   191101,
  \href{http://arxiv.org/abs/1406.7139}{{\ttfamily arXiv:1406.7139}}.

\bibitem{Lombriser2015c}
L.~{Lombriser} and A.~{Taylor}, {\em Journal of Cosmology and Astroparticle
  Physics} {\bf 3} (March 2016)   031,
  \href{http://arxiv.org/abs/1509.08458}{{\ttfamily arXiv:1509.08458}}.

\bibitem{Nishizawa2017}
A.~{Nishizawa}, {\em ArXiv e-prints}  (October 2017)
  \href{http://arxiv.org/abs/1710.04825}{{\ttfamily arXiv:1710.04825 [gr-qc]}}.

\bibitem{Chevallier2001}
M.~{Chevallier} and D.~{Polarski}, {\em International Journal of Modern Physics
  D} {\bf 10}  (2001) 213, \href{http://arxiv.org/abs/gr-qc/0009008}{{\ttfamily
  gr-qc/0009008}}.

\bibitem{Linder2003}
E.~V. {Linder}, {\em Physical Review Letters} {\bf 90} (March 2003)   091301,
  \href{http://arxiv.org/abs/astro-ph/0208512}{{\ttfamily astro-ph/0208512}}.

\bibitem{Lombriser2016a}
L.~{Lombriser} and N.~A. {Lima}, {\em Physics Letters B} {\bf 765} (February
  2017) 382, \href{http://arxiv.org/abs/1602.07670}{{\ttfamily
  arXiv:1602.07670}}.

\bibitem{Wetterich1988}
C.~{Wetterich}, {\em Nuclear Physics B} {\bf 302} (June 1988) 668,
  \href{http://arxiv.org/abs/1711.03844}{{\ttfamily arXiv:1711.03844
  [hep-th]}}.

\bibitem{Ratra:1988}
B.~{Ratra} and P.~J.~E. {Peebles}, {\em Physical Review D} {\bf 37} (June 1988)
  3406.

\bibitem{Crittenden2009}
R.~G. {Crittenden}, L.~{Pogosian} and G.-B. {Zhao}, {\em Journal of Cosmology
  and Astroparticle Physics} {\bf 12} (December 2009)   025,
  \href{http://arxiv.org/abs/astro-ph/0510293}{{\ttfamily astro-ph/0510293}}.

\bibitem{Brax2008}
P.~{Brax}, C.~{van de Bruck}, A.-C. {Davis} and D.~J. {Shaw}, {\em Physical
  Review D} {\bf 78} (November 2008)   104021,
  \href{http://arxiv.org/abs/0806.3415}{{\ttfamily arXiv:0806.3415}}.

\bibitem{Lombriser2013c}
L.~{Lombriser}, K.~{Koyama} and B.~{Li}, {\em Journal of Cosmology and
  Astroparticle Physics} {\bf 3} (March 2014)   021,
  \href{http://arxiv.org/abs/1312.1292}{{\ttfamily arXiv:1312.1292}}.

\bibitem{Ceron-Hurtado2016}
J.~J. Ceron-Hurtado, J.-h. He and B.~Li, {\em Phys. Rev.} {\bf D94}  (2016)
  064052, \href{http://arxiv.org/abs/1609.00532}{{\ttfamily arXiv:1609.00532
  [astro-ph.CO]}}.

\bibitem{Battye2017}
R.~A. {Battye}, B.~{Bolliet} and F.~{Pace}, {\em ArXiv e-prints}  (December
  2017) \href{http://arxiv.org/abs/1712.05976}{{\ttfamily arXiv:1712.05976}}.

\bibitem{ArmendarizPicon1999}
C.~{Armend{\'a}riz-Pic{\'o}n}, T.~{Damour} and V.~{Mukhanov}, {\em Physics
  Letters B} {\bf 458} (July 1999) 209,
  \href{http://arxiv.org/abs/hep-th/9904075}{{\ttfamily hep-th/9904075}}.

\bibitem{Deffayet2010}
C.~{Deffayet}, O.~{Pujol{\`a}s}, I.~{Sawicki} and A.~{Vikman}, {\em Journal of
  Cosmology and Astroparticle Physics} {\bf 10} (October 2010)   026,
  \href{http://arxiv.org/abs/1008.0048}{{\ttfamily arXiv:1008.0048 [hep-th]}}.

\bibitem{Nicolis2008}
A.~{Nicolis}, R.~{Rattazzi} and E.~{Trincherini}, {\em Physical Review D} {\bf
  79} (March 2009)   064036, \href{http://arxiv.org/abs/0811.2197}{{\ttfamily
  arXiv:0811.2197 [hep-th]}}.

\bibitem{Linder2015}
E.~V. {Linder}, G.~{Seng{\"o}r} and S.~{Watson}, {\em Journal of Cosmology and
  Astroparticle Physics} {\bf 5} (May 2016)   053,
  \href{http://arxiv.org/abs/1512.06180}{{\ttfamily arXiv:1512.06180}}.

\bibitem{Wang2012}
J.~{Wang}, L.~{Hui} and J.~{Khoury}, {\em Physical Review Letters} {\bf 109}
  (December 2012)   241301, \href{http://arxiv.org/abs/1208.4612}{{\ttfamily
  arXiv:1208.4612 [astro-ph.CO]}}.

\bibitem{Gleyzes2015}
J.~{Gleyzes}, D.~{Langlois}, M.~{Mancarella} and F.~{Vernizzi}, {\em Journal of
  Cosmology and Astroparticle Physics} {\bf 8} (August 2015)   054,
  \href{http://arxiv.org/abs/1504.05481}{{\ttfamily arXiv:1504.05481}}.

\bibitem{Horndeski1974}
G.~W. {Horndeski}, {\em International Journal of Theoretical Physics} {\bf 10}
  (September 1974) 363.

\bibitem{Lombriser2014}
L.~{Lombriser} and A.~{Taylor}, {\em Physical Review Letters} {\bf 114}
  (January 2015)   031101, \href{http://arxiv.org/abs/1405.2896}{{\ttfamily
  arXiv:1405.2896}}.

\bibitem{Abbott2017a}
B.~P. {Abbott}, R.~{Abbott}, T.~D. {Abbott}, F.~{Acernese}, K.~{Ackley},
  C.~{Adams}, T.~{Adams}, P.~{Addesso}, R.~X. {Adhikari}, V.~B. {Adya} and
  et~al., {\em Physical Review Letters} {\bf 119} (October 2017)   161101,
  \href{http://arxiv.org/abs/1710.05832}{{\ttfamily arXiv:1710.05832 [gr-qc]}}.

\bibitem{Abbott2017b}
B.~P. {Abbott}, R.~{Abbott}, T.~D. {Abbott}, F.~{Acernese}, K.~{Ackley},
  C.~{Adams}, T.~{Adams}, P.~{Addesso}, R.~X. {Adhikari}, V.~B. {Adya} and
  et~al., {\em Astrophysical Journal Letters} {\bf 848} (October 2017)   L13,
  \href{http://arxiv.org/abs/1710.05834}{{\ttfamily arXiv:1710.05834
  [astro-ph.HE]}}.

\bibitem{Nishizawa2014}
A.~{Nishizawa} and T.~{Nakamura}, {\em Physical Review D} {\bf 90} (August
  2014)   044048, \href{http://arxiv.org/abs/1406.5544}{{\ttfamily
  arXiv:1406.5544 [gr-qc]}}.

\bibitem{Ishak2005}
M.~{Ishak}, A.~{Upadhye} and D.~N. {Spergel}, {\em Physical Review D} {\bf 74}
  (August 2006)   043513,
  \href{http://arxiv.org/abs/astro-ph/0507184}{{\ttfamily astro-ph/0507184}}.

\bibitem{Mortonson2008}
M.~J. {Mortonson}, W.~{Hu} and D.~{Huterer}, {\em Physical Review D} {\bf 79}
  (January 2009)   023004, \href{http://arxiv.org/abs/0810.1744}{{\ttfamily
  arXiv:0810.1744}}.

\bibitem{Ruiz2014}
E.~J. {Ruiz} and D.~{Huterer}, {\em Physical Review D} {\bf 91} (March 2015)
  063009, \href{http://arxiv.org/abs/1410.5832}{{\ttfamily arXiv:1410.5832}}.

\bibitem{Lombriser2011a}
L.~{Lombriser}, {\em Physical Review D} {\bf 83} (March 2011)   063519,
  \href{http://arxiv.org/abs/1101.0594}{{\ttfamily arXiv:1101.0594}}.

\bibitem{Lombriser2013a}
L.~{Lombriser}, J.~{Yoo} and K.~{Koyama}, {\em Physical Review D} {\bf 87} (May
  2013)   104019, \href{http://arxiv.org/abs/1301.3132}{{\ttfamily
  arXiv:1301.3132 [astro-ph.CO]}}.

\bibitem{Lombriser2015b}
L.~{Lombriser} and A.~{Taylor}, {\em Journal of Cosmology and Astroparticle
  Physics} {\bf 11} (November 2015)   040,
  \href{http://arxiv.org/abs/1505.05915}{{\ttfamily arXiv:1505.05915}}.

\bibitem{Zhao2009}
G.-B. {Zhao}, L.~{Pogosian}, A.~{Silvestri} and J.~{Zylberberg}, {\em Physical
  Review D} {\bf 79} (April 2009)   083513,
  \href{http://arxiv.org/abs/0809.3791}{{\ttfamily arXiv:0809.3791}}.

\bibitem{Daniel2010}
S.~F. {Daniel}, E.~V. {Linder}, T.~L. {Smith}, R.~R. {Caldwell}, A.~{Cooray},
  A.~{Leauthaud} and L.~{Lombriser}, {\em Physical Review D} {\bf 81} (June
  2010)   123508, \href{http://arxiv.org/abs/1002.1962}{{\ttfamily
  arXiv:1002.1962}}.

\bibitem{Dossett2011}
J.~N. {Dossett}, M.~{Ishak} and J.~{Moldenhauer}, {\em Physical Review D} {\bf
  84} (December 2011)   123001,
  \href{http://arxiv.org/abs/1109.4583}{{\ttfamily arXiv:1109.4583}}.

\bibitem{Bonvin2018}
C.~{Bonvin} and P.~{Fleury}, {\em Journal of Cosmology and Astroparticle
  Physics} {\bf 5} (May 2018)   061,
  \href{http://arxiv.org/abs/1803.02771}{{\ttfamily arXiv:1803.02771}}.

\bibitem{Bertschinger2006}
E.~{Bertschinger}, {\em Astrophysical Journal} {\bf 648} (September 2006) 797,
  \href{http://arxiv.org/abs/astro-ph/0604485}{{\ttfamily astro-ph/0604485}}.

\bibitem{Battye2012}
R.~A. {Battye} and J.~A. {Pearson}, {\em Journal of Cosmology and Astroparticle
  Physics} {\bf 7} (July 2012)   019,
  \href{http://arxiv.org/abs/1203.0398}{{\ttfamily arXiv:1203.0398 [hep-th]}}.

\bibitem{Hu2014}
B.~{Hu}, M.~{Raveri}, N.~{Frusciante} and A.~{Silvestri}, {\em Physical Review
  D} {\bf 89} (May 2014)   103530,
  \href{http://arxiv.org/abs/1312.5742}{{\ttfamily arXiv:1312.5742}}.

\bibitem{Zumalacarregui2017}
M.~{Zumalac{\'a}rregui}, E.~{Bellini}, I.~{Sawicki}, J.~{Lesgourgues} and P.~G.
  {Ferreira}, {\em Journal of Cosmology and Astroparticle Physics} {\bf 8}
  (August 2017)   019, \href{http://arxiv.org/abs/1605.06102}{{\ttfamily
  arXiv:1605.06102}}.

\bibitem{Bloomfield2013}
J.~{Bloomfield}, {\em Journal of Cosmology and Astroparticle Physics} {\bf 12}
  (December 2013)   044, \href{http://arxiv.org/abs/1304.6712}{{\ttfamily
  arXiv:1304.6712 [astro-ph.CO]}}.

\bibitem{Horava2009}
P.~{Ho{\v r}ava}, {\em Physical Review D} {\bf 79} (April 2009)   084008,
  \href{http://arxiv.org/abs/0901.3775}{{\ttfamily arXiv:0901.3775 [hep-th]}}.

\bibitem{Moore2001}
G.~D. {Moore} and A.~E. {Nelson}, {\em Journal of High Energy Physics} {\bf 9}
  (September 2001)   023, \href{http://arxiv.org/abs/hep-ph/0106220}{{\ttfamily
  hep-ph/0106220}}.

\bibitem{Caves1980}
C.~M. {Caves}, {\em Annals of Physics} {\bf 125} (March 1980) 35.

\bibitem{Jimenez2015}
J.~{Beltr{\'a}n Jim{\'e}nez}, F.~{Piazza} and H.~{Velten}, {\em Physical Review
  Letters} {\bf 116} (February 2016)   061101,
  \href{http://arxiv.org/abs/1507.05047}{{\ttfamily arXiv:1507.05047 [gr-qc]}}.

\bibitem{Battye2018}
R.~A. {Battye}, F.~{Pace} and D.~{Trinh}, {\em ArXiv e-prints}  (February 2018)
  \href{http://arxiv.org/abs/1802.09447}{{\ttfamily arXiv:1802.09447}}.

\bibitem{Brax2015}
P.~{Brax}, C.~{Burrage} and A.-C. {Davis}, {\em Journal of Cosmology and
  Astroparticle Physics} {\bf 3} (March 2016)   004,
  \href{http://arxiv.org/abs/1510.03701}{{\ttfamily arXiv:1510.03701 [gr-qc]}}.

\bibitem{Barreira2014b}
A.~{Barreira}, B.~{Li}, C.~M. {Baugh} and S.~{Pascoli}, {\em Journal of
  Cosmology and Astroparticle Physics} {\bf 8} (August 2014)   059,
  \href{http://arxiv.org/abs/1406.0485}{{\ttfamily arXiv:1406.0485}}.

\bibitem{Lombriser2009}
L.~{Lombriser}, W.~{Hu}, W.~{Fang} and U.~{Seljak}, {\em Physical Review D}
  {\bf 80} (September 2009)   063536,
  \href{http://arxiv.org/abs/0905.1112}{{\ttfamily arXiv:0905.1112
  [astro-ph.CO]}}.

\bibitem{Kimura2011}
R.~{Kimura}, T.~{Kobayashi} and K.~{Yamamoto}, {\em Physical Review D} {\bf 85}
  (June 2012)   123503, \href{http://arxiv.org/abs/1110.3598}{{\ttfamily
  arXiv:1110.3598 [astro-ph.CO]}}.

\bibitem{Kimura2011b}
R.~{Kimura} and K.~{Yamamoto}, {\em Journal of Cosmology and Astroparticle
  Physics} {\bf 7} (July 2012)   050,
  \href{http://arxiv.org/abs/1112.4284}{{\ttfamily arXiv:1112.4284
  [astro-ph.CO]}}.

\bibitem{McManus2016}
R.~{McManus}, L.~{Lombriser} and J.~{Pe{\~n}arrubia}, {\em Journal of Cosmology
  and Astroparticle Physics} {\bf 11} (November 2016)   006,
  \href{http://arxiv.org/abs/1606.03282}{{\ttfamily arXiv:1606.03282 [gr-qc]}}.

\bibitem{Cornish2017}
N.~{Cornish}, D.~{Blas} and G.~{Nardini}, {\em Physical Review Letters} {\bf
  119} (October 2017)   161102,
  \href{http://arxiv.org/abs/1707.06101}{{\ttfamily arXiv:1707.06101 [gr-qc]}}.

\bibitem{Blas2016}
D.~{Blas}, M.~M. {Ivanov}, I.~{Sawicki} and S.~{Sibiryakov}, {\em Soviet
  Journal of Experimental and Theoretical Physics Letters} {\bf 103} (May 2016)
  624, \href{http://arxiv.org/abs/1602.04188}{{\ttfamily arXiv:1602.04188
  [gr-qc]}}.

\bibitem{Bettoni2016}
D.~{Bettoni}, J.~M. {Ezquiaga}, K.~{Hinterbichler} and M.~{Zumalac{\'a}rregui},
  {\em Physical Review D} {\bf 95} (April 2017)   084029,
  \href{http://arxiv.org/abs/1608.01982}{{\ttfamily arXiv:1608.01982 [gr-qc]}}.

\bibitem{Raveri2014}
M.~{Raveri}, C.~{Baccigalupi}, A.~{Silvestri} and S.-Y. {Zhou}, {\em Physical
  Review D} {\bf 91} (March 2015)   061501,
  \href{http://arxiv.org/abs/1405.7974}{{\ttfamily arXiv:1405.7974}}.

\bibitem{Amendola2014}
L.~{Amendola}, G.~{Ballesteros} and V.~{Pettorino}, {\em Physical Review D}
  {\bf 90} (August 2014)   043009,
  \href{http://arxiv.org/abs/1405.7004}{{\ttfamily arXiv:1405.7004}}.

\bibitem{Schutz1986}
B.~F. {Schutz}, {\em Nature} {\bf 323} (September 1986)   310.

\bibitem{Holz2005}
D.~E. {Holz} and S.~A. {Hughes}, {\em Astrophysical Journal} {\bf 629} (August
  2005) 15, \href{http://arxiv.org/abs/astro-ph/0504616}{{\ttfamily
  astro-ph/0504616}}.

\bibitem{Belgacem2017}
E.~{Belgacem}, Y.~{Dirian}, S.~{Foffa} and M.~{Maggiore}, {\em ArXiv e-prints}
  (December 2017) \href{http://arxiv.org/abs/1712.08108}{{\ttfamily
  arXiv:1712.08108}}.

\bibitem{Amendola2017}
L.~{Amendola}, I.~{Sawicki}, M.~{Kunz} and I.~D. {Saltas}, {\em ArXiv e-prints}
   (December 2017) \href{http://arxiv.org/abs/1712.08623}{{\ttfamily
  arXiv:1712.08623}}.

\bibitem{Abbott2017c}
B.~P. {Abbott}, R.~{Abbott}, T.~D. {Abbott}, F.~{Acernese}, K.~{Ackley},
  C.~{Adams}, T.~{Adams}, P.~{Addesso}, R.~X. {Adhikari}, V.~B. {Adya} and
  et~al., {\em Nature} {\bf 551} (November 2017) 85,
  \href{http://arxiv.org/abs/1710.05835}{{\ttfamily arXiv:1710.05835}}.

\bibitem{MacLeod2007}
C.~L. {MacLeod} and C.~J. {Hogan}, {\em Physical Review D} {\bf 77} (February
  2008)   043512, \href{http://arxiv.org/abs/0712.0618}{{\ttfamily
  arXiv:0712.0618}}.

\bibitem{Tamanini2016}
N.~{Tamanini}, C.~{Caprini}, E.~{Barausse}, A.~{Sesana}, A.~{Klein} and
  A.~{Petiteau}, {\em Journal of Cosmology and Astroparticle Physics} {\bf 4}
  (April 2016)   002, \href{http://arxiv.org/abs/1601.07112}{{\ttfamily
  arXiv:1601.07112}}.

\bibitem{Silvestri2013}
A.~{Silvestri}, L.~{Pogosian} and R.~V. {Buniy}, {\em Physical Review D} {\bf
  87} (May 2013)   104015, \href{http://arxiv.org/abs/1302.1193}{{\ttfamily
  arXiv:1302.1193}}.

\bibitem{Bertschinger2008}
E.~{Bertschinger} and P.~{Zukin}, {\em Physical Review D} {\bf 78} (July 2008)
   024015, \href{http://arxiv.org/abs/0801.2431}{{\ttfamily arXiv:0801.2431}}.

\bibitem{Simpson2012}
F.~{Simpson}, C.~{Heymans}, D.~{Parkinson}, C.~{Blake}, M.~{Kilbinger},
  J.~{Benjamin}, T.~{Erben}, H.~{Hildebrandt}, H.~{Hoekstra}, T.~D. {Kitching},
  Y.~{Mellier}, L.~{Miller}, L.~{Van Waerbeke}, J.~{Coupon}, L.~{Fu},
  J.~{Harnois-D{\'e}raps}, M.~J. {Hudson}, K.~{Kuijken}, B.~{Rowe},
  T.~{Schrabback}, E.~{Semboloni}, S.~{Vafaei} and M.~{Velander}, {\em Monthly
  Notices of the Royal Astronomical Society} {\bf 429} (March 2013) 2249,
  \href{http://arxiv.org/abs/1212.3339}{{\ttfamily arXiv:1212.3339}}.

\bibitem{Lima2016}
N.~A. {Lima}, V.~{Smer-Barreto} and L.~{Lombriser}, {\em Physical Review D}
  {\bf 94} (October 2016)   083507,
  \href{http://arxiv.org/abs/1603.05239}{{\ttfamily arXiv:1603.05239}}.

\bibitem{Kennedy2018}
J.~Kennedy, L.~Lombriser and A.~Taylor  (2018)
  \href{http://arxiv.org/abs/1804.04582}{{\ttfamily arXiv:1804.04582
  [astro-ph.CO]}}.

\bibitem{Laureijs2011}
R.~{Laureijs}, J.~{Amiaux}, S.~{Arduini}, J.~. {Augu{\`e}res}, J.~{Brinchmann},
  R.~{Cole}, M.~{Cropper}, C.~{Dabin}, L.~{Duvet}, A.~{Ealet} and et~al., {\em
  ArXiv e-prints}  (October 2011)
  \href{http://arxiv.org/abs/1110.3193}{{\ttfamily arXiv:1110.3193
  [astro-ph.CO]}}.

\bibitem{Ivezic2008}
{\v Z}.~{Ivezi{\'c}}, S.~M. {Kahn}, J.~A. {Tyson}, B.~{Abel}, E.~{Acosta},
  R.~{Allsman}, D.~{Alonso}, Y.~{AlSayyad}, S.~F. {Anderson}, J.~{Andrew} and
  et~al., {\em ArXiv e-prints}  (May 2008)
  \href{http://arxiv.org/abs/0805.2366}{{\ttfamily arXiv:0805.2366}}.

\bibitem{Dvali2000}
G.~{Dvali}, G.~{Gabadadze} and M.~{Porrati}, {\em Physics Letters B} {\bf 485}
  (July 2000) 208, \href{http://arxiv.org/abs/hep-th/0005016}{{\ttfamily
  hep-th/0005016}}.

\bibitem{Buchdahl1970}
H.~A. {Buchdahl}, {\em Monthly Notices of the Royal Astronomical Society} {\bf
  150}  (1970)  ~1.

\bibitem{Brax2013}
P.~{Brax} and P.~{Valageas}, {\em Physical Review D} {\bf 88} (July 2013)
  023527, \href{http://arxiv.org/abs/1305.5647}{{\ttfamily arXiv:1305.5647}}.

\bibitem{Lombriser2014a}
L.~{Lombriser}, {\em Annalen der Physik} {\bf 526} (August 2014) 259,
  \href{http://arxiv.org/abs/1403.4268}{{\ttfamily arXiv:1403.4268}}.

\bibitem{Schmidt2008}
F.~{Schmidt}, M.~{Lima}, H.~{Oyaizu} and W.~{Hu}, {\em Physical Review D} {\bf
  79} (April 2009)   083518, \href{http://arxiv.org/abs/0812.0545}{{\ttfamily
  arXiv:0812.0545}}.

\bibitem{Pace2010}
F.~{Pace}, J.-C. {Waizmann} and M.~{Bartelmann}, {\em Monthly Notices of the
  Royal Astronomical Society} {\bf 406} (August 2010) 1865,
  \href{http://arxiv.org/abs/1005.0233}{{\ttfamily arXiv:1005.0233}}.

\bibitem{Borisov2011}
A.~{Borisov}, B.~{Jain} and P.~{Zhang}, {\em Physical Review D} {\bf 85} (March
  2012)   063518, \href{http://arxiv.org/abs/1102.4839}{{\ttfamily
  arXiv:1102.4839}}.

\bibitem{Kopp2013}
M.~{Kopp}, S.~A. {Appleby}, I.~{Achitouv} and J.~{Weller}, {\em Physical Review
  D} {\bf 88} (October 2013)   084015,
  \href{http://arxiv.org/abs/1306.3233}{{\ttfamily arXiv:1306.3233
  [astro-ph.CO]}}.

\bibitem{Li2011c}
B.~{Li} and G.~{Efstathiou}, {\em Monthly Notices of the Royal Astronomical
  Society} {\bf 421} (April 2012) 1431,
  \href{http://arxiv.org/abs/1110.6440}{{\ttfamily arXiv:1110.6440
  [astro-ph.CO]}}.

\bibitem{Li2012}
B.~{Li} and T.~Y. {Lam}, {\em Monthly Notices of the Royal Astronomical
  Society} {\bf 425} (September 2012) 730,
  \href{http://arxiv.org/abs/1205.0058}{{\ttfamily arXiv:1205.0058
  [astro-ph.CO]}}.

\bibitem{Lombriser2013b}
L.~{Lombriser}, B.~{Li}, K.~{Koyama} and G.-B. {Zhao}, {\em Physical Review D}
  {\bf 87} (June 2013)   123511,
  \href{http://arxiv.org/abs/1304.6395}{{\ttfamily arXiv:1304.6395
  [astro-ph.CO]}}.

\bibitem{Lombriser2012}
L.~{Lombriser}, K.~{Koyama}, G.-B. {Zhao} and B.~{Li}, {\em Physical Review D}
  {\bf 85} (June 2012)   124054,
  \href{http://arxiv.org/abs/1203.5125}{{\ttfamily arXiv:1203.5125
  [astro-ph.CO]}}.

\bibitem{Lombriser2011b}
L.~{Lombriser}, F.~{Schmidt}, T.~{Baldauf}, R.~{Mandelbaum}, U.~{Seljak} and
  R.~E. {Smith}, {\em Physical Review D} {\bf 85} (May 2012)   102001,
  \href{http://arxiv.org/abs/1111.2020}{{\ttfamily arXiv:1111.2020
  [astro-ph.CO]}}.

\bibitem{Gronke2015}
M.~{Gronke}, D.~F. {Mota} and H.~A. {Winther}, {\em Astronomy \& Astrophysics}
  {\bf 583} (November 2015)   A123,
  \href{http://arxiv.org/abs/1505.07129}{{\ttfamily arXiv:1505.07129}}.

\bibitem{Mitchell2018}
M.~A. {Mitchell}, J.-h. {He}, C.~{Arnold} and B.~{Li}, {\em ArXiv e-prints}
  (February 2018) \href{http://arxiv.org/abs/1802.02165}{{\ttfamily
  arXiv:1802.02165}}.

\bibitem{Schmidt2009}
F.~{Schmidt}, W.~{Hu} and M.~{Lima}, {\em Physical Review D} {\bf 81} (March
  2010)   063005, \href{http://arxiv.org/abs/0911.5178}{{\ttfamily
  arXiv:0911.5178 [astro-ph.CO]}}.

\bibitem{Lombriser2010}
L.~{Lombriser}, A.~{Slosar}, U.~{Seljak} and W.~{Hu}, {\em Physical Review D}
  {\bf 85} (June 2012)   124038,
  \href{http://arxiv.org/abs/1003.3009}{{\ttfamily arXiv:1003.3009}}.

\bibitem{Barreira2014a}
A.~{Barreira}, B.~{Li}, W.~A. {Hellwing}, L.~{Lombriser}, C.~M. {Baugh} and
  S.~{Pascoli}, {\em Journal of Cosmology and Astroparticle Physics} {\bf 4}
  (April 2014)   029, \href{http://arxiv.org/abs/1401.1497}{{\ttfamily
  arXiv:1401.1497}}.

\bibitem{Cataneo2016}
M.~Cataneo, D.~Rapetti, L.~Lombriser and B.~Li, {\em JCAP} {\bf 1612}  (2016)
  024, \href{http://arxiv.org/abs/1607.08788}{{\ttfamily arXiv:1607.08788
  [astro-ph.CO]}}.

\bibitem{Hagstotz2018}
S.~{Hagstotz}, M.~{Costanzi}, M.~{Baldi} and J.~{Weller}, {\em ArXiv e-prints}
  (June 2018) \href{http://arxiv.org/abs/1806.07400}{{\ttfamily
  arXiv:1806.07400}}.

\bibitem{Clampitt2012}
J.~{Clampitt}, Y.-C. {Cai} and B.~{Li}, {\em Monthly Notices of the Royal
  Astronomical Society} {\bf 431} (May 2013) 749,
  \href{http://arxiv.org/abs/1212.2216}{{\ttfamily arXiv:1212.2216
  [astro-ph.CO]}}.

\bibitem{Lam2014}
T.~Y. {Lam}, J.~{Clampitt}, Y.-C. {Cai} and B.~{Li}, {\em Monthly Notices of
  the Royal Astronomical Society} {\bf 450} (July 2015) 3319,
  \href{http://arxiv.org/abs/1408.5338}{{\ttfamily arXiv:1408.5338}}.

\bibitem{Peacock2000}
J.~A. {Peacock} and R.~E. {Smith}, {\em Monthly Notices of the Royal
  Astronomical Society} {\bf 318} (November 2000) 1144,
  \href{http://arxiv.org/abs/astro-ph/0005010}{{\ttfamily astro-ph/0005010}}.

\bibitem{Seljak2000}
U.~{Seljak}, {\em Monthly Notices of the Royal Astronomical Society} {\bf 318}
  (October 2000) 203, \href{http://arxiv.org/abs/astro-ph/0001493}{{\ttfamily
  astro-ph/0001493}}.

\bibitem{Cooray2002}
A.~{Cooray} and R.~{Sheth}, {\em Physics Reports} {\bf 372} (December 2002) 1,
  \href{http://arxiv.org/abs/astro-ph/0206508}{{\ttfamily astro-ph/0206508}}.

\bibitem{Achitouv2015}
I.~{Achitouv}, M.~{Baldi}, E.~{Puchwein} and J.~{Weller}, {\em Physical Review
  D} {\bf 93} (May 2016)   103522,
  \href{http://arxiv.org/abs/1511.01494}{{\ttfamily arXiv:1511.01494}}.

\bibitem{Li2011a}
Y.~{Li} and W.~{Hu}, {\em Physical Review D} {\bf 84} (October 2011)   084033,
  \href{http://arxiv.org/abs/1107.5120}{{\ttfamily arXiv:1107.5120}}.

\bibitem{Mead2016}
A.~J. {Mead}, C.~{Heymans}, L.~{Lombriser}, J.~A. {Peacock}, O.~I. {Steele} and
  H.~A. {Winther}, {\em Monthly Notices of the Royal Astronomical Society} {\bf
  459} (June 2016) 1468, \href{http://arxiv.org/abs/1602.02154}{{\ttfamily
  arXiv:1602.02154}}.

\bibitem{Fedeli2014}
C.~{Fedeli}, {\em Journal of Cosmology and Astroparticle Physics} {\bf 4}
  (April 2014)   028, \href{http://arxiv.org/abs/1401.2997}{{\ttfamily
  arXiv:1401.2997}}.

\bibitem{Terukina2013}
A.~{Terukina}, L.~{Lombriser}, K.~{Yamamoto}, D.~{Bacon}, K.~{Koyama} and R.~C.
  {Nichol}, {\em Journal of Cosmology and Astroparticle Physics} {\bf 4} (April
  2014)   013, \href{http://arxiv.org/abs/1312.5083}{{\ttfamily
  arXiv:1312.5083}}.

\bibitem{Sakstein2016}
J.~{Sakstein}, H.~{Wilcox}, D.~{Bacon}, K.~{Koyama} and R.~C. {Nichol}, {\em
  Journal of Cosmology and Astroparticle Physics} {\bf 7} (July 2016)   019,
  \href{http://arxiv.org/abs/1603.06368}{{\ttfamily arXiv:1603.06368}}.

\bibitem{Lombriser2015a}
L.~{Lombriser}, F.~{Simpson} and A.~{Mead}, {\em Physical Review Letters} {\bf
  114} (June 2015)   251101, \href{http://arxiv.org/abs/1501.04961}{{\ttfamily
  arXiv:1501.04961}}.

\bibitem{Winther2014}
H.~A. {Winther} and P.~G. {Ferreira}, {\em Physical Review D} {\bf 91} (June
  2015)   123507, \href{http://arxiv.org/abs/1403.6492}{{\ttfamily
  arXiv:1403.6492}}.

\bibitem{Mead2014}
A.~J. {Mead}, J.~A. {Peacock}, L.~{Lombriser} and B.~{Li}, {\em Monthly Notices
  of the Royal Astronomical Society} {\bf 452} (October 2015) 4203,
  \href{http://arxiv.org/abs/1412.5195}{{\ttfamily arXiv:1412.5195}}.

\bibitem{Bertotti2003}
B.~{Bertotti}, L.~{Iess} and P.~{Tortora}, {\em Nature} {\bf 425} (September
  2003) 374.

\bibitem{Zhang2016}
X.~{Zhang}, W.~{Zhao}, H.~{Huang} and Y.~{Cai}, {\em Physical Review D} {\bf
  93} (June 2016)   124003, \href{http://arxiv.org/abs/1603.09450}{{\ttfamily
  arXiv:1603.09450 [gr-qc]}}.

\bibitem{Gabadadze2012}
G.~{Gabadadze}, K.~{Hinterbichler} and D.~{Pirtskhalava}, {\em Physical Review
  D} {\bf 85} (June 2012)   125007,
  \href{http://arxiv.org/abs/1202.6364}{{\ttfamily arXiv:1202.6364 [hep-th]}}.

\bibitem{Padilla2012}
A.~{Padilla} and P.~M. {Saffin}, {\em Journal of High Energy Physics} {\bf 7}
  (July 2012)   122, \href{http://arxiv.org/abs/1204.1352}{{\ttfamily
  arXiv:1204.1352 [hep-th]}}.

\bibitem{Epstein1975}
R.~{Epstein} and R.~V. {Wagoner}, {\em The Astrophysical Journal} {\bf 197}
  (May 1975) 717.

\bibitem{Will1996}
C.~M. {Will} and A.~G. {Wiseman}, {\em Physical Review D} {\bf 54} (October
  1996) 4813, \href{http://arxiv.org/abs/gr-qc/9608012}{{\ttfamily
  gr-qc/9608012}}.

\bibitem{Pati2002}
M.~E. {Pati} and C.~M. {Will}, {\em Physical Review D} {\bf 65} (May 2002)
  104008, \href{http://arxiv.org/abs/gr-qc/0201001}{{\ttfamily gr-qc/0201001}}.

\bibitem{Thomas2015}
D.~B. {Thomas}, M.~{Bruni} and D.~{Wands}, {\em Journal of Cosmology and
  Astroparticle Physics} {\bf 9} (September 2015)   021,
  \href{http://arxiv.org/abs/1403.4947}{{\ttfamily arXiv:1403.4947}}.

\bibitem{Kase2014}
R.~{Kase}, L.~{\'A}. {Gergely} and S.~{Tsujikawa}, {\em Physical Review D} {\bf
  90} (December 2014)   124019,
  \href{http://arxiv.org/abs/1406.2402}{{\ttfamily arXiv:1406.2402 [hep-th]}}.

\bibitem{Brax2012b}
P.~{Brax}, A.-C. {Davis}, B.~{Li}, H.~A. {Winther} and G.-B. {Zhao}, {\em
  Journal of Cosmology and Astroparticle Physics} {\bf 10} (October 2012)
  002, \href{http://arxiv.org/abs/1206.3568}{{\ttfamily arXiv:1206.3568}}.

\bibitem{Kennedy2017}
J.~{Kennedy}, L.~{Lombriser} and A.~{Taylor}, {\em Physical Review D} {\bf 96}
  (October 2017)   084051, \href{http://arxiv.org/abs/1705.09290}{{\ttfamily
  arXiv:1705.09290 [gr-qc]}}.

\end{thebibliography}


\begin{thebibliography}{}

\end{thebibliography}

\end{document}